\documentclass[a4paper,twocolumn,10pt,accepted=2024-09-19]{quantumarticle}

\pdfoutput=1 
\usepackage{CJK}

\usepackage{tikzit}

\tikzstyle{state}=[fill=white, draw=black, shape=isosceles triangle, isosceles triangle stretches=false, inner sep=0.2, shape border rotate=-90, isosceles triangle apex angle=100, minimum width=2 cm, line width=0.025cm]
\tikzstyle{effect}=[fill=white, draw=black, shape=isosceles triangle, isosceles triangle stretches=true, inner sep=0.02cm, shape border rotate=90, minimum width=1, line width=0.025cm, isosceles triangle apex angle=100]
\tikzstyle{discard}=[shape=tlground, fill=white, draw=black, rotate=180, scale=1.2]
\tikzstyle{bistate}=[fill=white, draw=black, shape=isosceles triangle, isosceles triangle stretches=true, inner sep=0, shape border rotate=-90, isosceles triangle apex angle=95, minimum width=1cm, line width=0.025cm]
\tikzstyle{identity}=[shape=tlground, fill=white, draw=black, scale=1.2]
\tikzstyle{SAdiscard}=[shape=ground, fill=white, draw=black, rotate=180]
\tikzstyle{SAidentity}=[shape=ground, fill=white, draw=black]
\tikzstyle{diamond}=[fill=white, draw=black, shape=diamond, minimum width=1cm, inner sep=0, minimum height=1cm, line width=0.025cm]

\tikzstyle{black}=[-, draw=black, line width=0.025cm]
\tikzstyle{da}=[-, dashed, gray]
\tikzstyle{arrow}=[->, draw=black, line width=0.025cm]
\tikzstyle{gray}=[-, draw=gray]
\tikzstyle{doublearrow}=[<->, line width=0.025cm, draw=black]

\usepackage{mathtools,amssymb,amsfonts,amsthm,physics,subcaption,graphicx,multirow,float,bm}
\usepackage[colorlinks,citecolor=blue,linkcolor=blue,urlcolor=blue, breaklinks=true]{hyperref}
\usepackage[shortlabels]{enumitem}
\counterwithout{figure}{section}
\usepackage[none]{hyphenat}
\usepackage{csquotes}

\usepackage{color}

\setlist{nosep}

\usepackage{thmtools} 
\usepackage{thm-restate}

\newtheorem{theorem}{Theorem}
\newtheorem{corollary}{Corollary}
\newtheorem{definition}{Definition}
\newcommand{\obs}{\mathtt{obs}}

\newcommand{\pa}{\mathtt{pa}}

\newcommand{\opa}{\mathtt{opa}}
\newcommand{\Opa}{\mathtt{Opa}}

\let\oldperp\perp
\renewcommand{\perp}{\mathbin{\oldperp_{\text{d}}}}

\newcommand{\RCA}{{Relativistic Causal Arrow}}

\newcommand{\AOE}{{Absoluteness of Observed Events}}

\newcommand{\IS}{{Independent Settings}}

\newcommand{\LA}{{Local Agency}}

\begin{document}

\title{Relating Wigner's Friend Scenarios to Nonclassical Causal Compatibility, Monogamy Relations, and Fine Tuning}

\begin{CJK*}{UTF8}{gbsn}

\author{Y{\`i}l{\`e} Y{\=\i}ng 
} 
\email{yying@pitp.ca}
\author{Marina Maciel Ansanelli}
\affiliation{Perimeter Institute for Theoretical Physics, Waterloo, Ontario, Canada, N2L 2Y5}
\affiliation{Department of Physics and Astronomy, University of Waterloo, Waterloo, Ontario, Canada, N2L 3G1}
\author{Andrea Di Biagio}
\affiliation{Institute for Quantum Optics and Quantum Information (IQOQI) Vienna, Austrian Academy of Sciences, Boltzmanngasse 3, A-1090 Vienna, Austria}
\affiliation{Basic Research Community for Physics e.V., Mariannenstraße 89, Leipzig, Germany}
\author{Elie Wolfe}
\affiliation{Perimeter Institute for Theoretical Physics, Waterloo, Ontario, Canada, N2L 2Y5}
\affiliation{Department of Physics and Astronomy, University of Waterloo, Waterloo, Ontario, Canada, N2L 3G1}
\author{David Schmid}
\affiliation{International Centre for Theory of Quantum Technologies, University of Gda{\'n}sk, 80-309 Gda\'nsk, Poland}
\author{Eric Gama Cavalcanti}
\affiliation{Centre for Quantum Dynamics, Griffith University, Yugambeh Country, 
Gold Coast, Queensland 4222, Australia}

\begin{abstract}

Nonclassical causal modeling was developed in order to explain violations of Bell inequalities while adhering to relativistic causal structure and \emph{faithfulness}---that is, avoiding fine-tuned causal explanations. Recently, a no-go theorem  that can be viewed as being stronger than Bell's theorem has been derived, based on extensions of the Wigner's friend thought experiment: the Local Friendliness (LF) no-go theorem.
Here we show that the LF no-go theorem poses formidable challenges for the field of causal modeling, even when nonclassical and/or cyclic causal explanations are considered.
We first recast the LF inequalities, one of the key elements of the LF no-go theorem, as special cases of monogamy relations stemming from a statistical marginal problem. We then further recast LF inequalities as causal compatibility inequalities stemming from a \emph{nonclassical} causal marginal problem, for a causal structure implied by well-motivated causal-metaphysical assumptions.  We find that the LF inequalities emerge from this causal structure even when one allows the latent causes of observed events to admit post-quantum descriptions, such as in a generalized probabilistic theory or in an even more exotic theory. We further prove that \emph{no} nonclassical causal model can explain violations of LF inequalities without violating the No Fine-Tuning principle. 
Finally, we note that these obstacles cannot be overcome even if one appeals to \emph{cyclic} causal models, and we discuss potential directions for further extensions of the causal modeling framework. 
\end{abstract}

\maketitle

\end{CJK*}
\tableofcontents

\raggedbottom

\section{Introduction}
\label{sec_intro}
The dynamic synergy between the field of quantum foundations and the field of causal inference can be traced back to as early as \cite{glymourMarkov2006} and gained substantial momentum through the groundbreaking contributions of \cite{woodLesson2015}. This influential work brought forth a novel perspective on Bell's theorem, one of the most pivotal results in quantum foundations, by utilizing Pearl's framework of classical causal models~\cite{causality_pearl}.

In this framework, causal structures are represented using Directed Acyclic Graphs (DAGs). For a classical causal model, a node in such a graph represents a random variable associated with either an observed event (such a node is called an \emph{observed} node) or some other event or system that is useful for explaining the correlations between those observed events (such a node is called a \emph{latent} node). An arrow in a DAG represents a direct cause-effect relationship. 

\begin{figure}[!h]
    \centering     \includegraphics[width=0.2\textwidth]{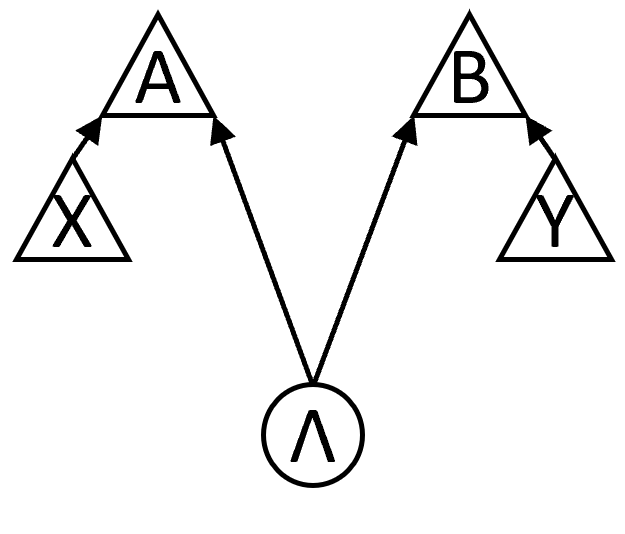}
    \caption{\textbf{The Bell DAG.} Triangles represent observed nodes while circles represent latent nodes. }
    \label{fig_Bell}
\end{figure}

As an example, one can encode the causal assumptions that give rise to Bell's theorem in the DAG of Fig.~\ref{fig_Bell}, known as the \emph{Bell DAG}. There,  $X$, $Y$, $A$, and $B$ are observed while $\Lambda$ is latent. Specifically, $X$ and $Y$ represent Alice's and Bob's measurement settings, and $A$ and $B$ represent their respective outcomes, while $\Lambda$ represents a complete specification of all common causes between $A$ and $B$~\cite{Bell_1964}. 

This DAG is motivated by causal intuitions regarding the Bell experiment, and can be rigorously derived via two independent avenues: causal-metaphysical assumptions such as Local Causality~\cite{Bell_1976}, or by causal discovery principles such as No Fine-Tuning (or faithfulness)~\cite{woodLesson2015,cavalcantiClassical2018,Pearl_2021}. No Fine-Tuning is the assumption that observed statistical independences ought to be explained by the underlying causal structure rather than by fine-tuning the quantitative causal relationships among the nodes.

A causal structure imposes constraints on the compatible probability distributions over its observed nodes, known as \textit{causal compatibility constraints}. The causal compatibility constraints can take the form of equalities or inequalities, which we respectively call \textit{causal compatibility equalities or inequalities}. For example, when the latent node $\Lambda$ of the Bell DAG is associated with a classical random variable, Bell inequalities are causal compatibility inequalities that follow from this DAG. The central message of Bell's theorem can then be understood as the inability of the Bell DAG to explain violations of Bell inequalities under classical causal modeling~\cite{woodLesson2015}.

Ref.~\cite{woodLesson2015}  
established a bridge between quantum physicists and the classical causal inference community, paving the way for fresh insights on both fronts. For instance, it underscored the significance of causal compatibility inequalities (as opposed to equalities) for statisticians, and encouraged physicists to explore quantum advantages in more general causal structures~\cite{Fritz_2012, Triangle_TC, bipartipe_CoiteuxRoy, Lauand2023, experimental_instrumental, Experimental_Triangle} and to contribute to classical causal inference~\cite{Wolfe_inflation,Navascues_inflation_complete,Rosset_bound,Fraser_Combinatorial_Solution,Zjawin_restricted,Berkeley_Workshop,PI_Workshop}. 

Moreover, the idea of providing a causal explanation for violations of Bell inequalities while keeping the Bell causal structure intact (i.e., without adding any extra arrows) spurred the development of nonclassical generalizations to Pearl's classical causal model framework, where one permits correlations to be explained by causes whose properties are described by quantum theory or other generalized probabilistic theories (GPTs) ~\cite{hensonTheoryindependent2014,barrett2020quantum,AllenQuantumCommonCauses2017,costaQuantum2016}. Substantial progress in nonclassical causal inference has unfolded since then~\cite{quantum_inflation,Weilenmann2020analysingcausal,multipartite_nonlocal,chaves_informationtheoretic_2015}.

Independent of the above considerations, another way in which the classical causal models framework has been generalized is to allow cyclic instead of only acyclic causal structures. Cyclic causal structures have been used in classical causal inference for situations with feedback loops~\cite{forre2017markov,Bongers2021} and have been studied in the quantum context concerning indefinite causal order~\cite{Araujo2017,Barrett_2021,Oreshkov_2012}. 

In this paper, we use a framework for nonclassical causal models that is built upon the work of~\cite{hensonTheoryindependent2014}, a widely adopted framework for nonclassical causal inference. We also extend this framework to include cyclic causal models (and then prove our results within this more general context).

The present work aims to construct a similar bridge to the one built by Ref.~\cite{woodLesson2015}, but where we connect extended Wigner's Friend no-go theorems~\cite{schmid2023review} with \emph{nonclassical} causal inference.  We achieve this by demonstrating that the Local Friendliness (LF) inequalities~\cite{bongStrong2020}, featured in one of the most significant extended Wigner's Friend no-go theorems, can be viewed as \emph{nonclassical} causal compatibility inequalities for a causal structure (the LF DAG, see Fig.~\ref{fig_LF}) implied by well-motivated causal-metaphysical assumptions underlying the Local Friendliness assumptions~\cite{bongStrong2020}.  Consequently, we prove that these inequalities need to be respected even if one allows quantum, GPT, or other more unconventional nonclassical causal explanations. 
Moreover, we demonstrate that these causal compatibility inequalities can be derived from the requirement of No Fine-Tuning. Importantly, our No Fine-Tuning argument can also be applied to \textit{cyclic} causal models.

\subsection{The minimal LF scenario}
\label{sec_LFscenario}

\begin{figure}
    \centering
    \includegraphics{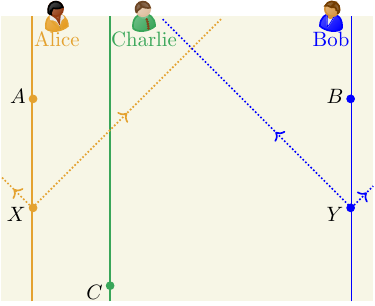}
    \caption{The minimal Local Friendliness (LF) scenario with additional spatio-temporal requirements. The solid lines are the worldlines of the three observers. The dotted lines indicate light cones. Note that such spatio-temporal requirements are not needed for any of our results except those in Sec.~\ref{sec_RCA}. This figure is adapted from Fig.1 of Ref.~\cite{wisemanThoughtful2022}.}
    \label{fig_spacetime}
\end{figure}

The specific scenario we consider in this work is the minimal\footnote{All of our results should be generalizable to LF scenarios with two or more pairs of observers and superobservers, such as that in Ref.~\cite{bongStrong2020} (where Bob also has a corresponding friend, Debbie). Here we use the minimal LF scenario for simplicity.} LF scenario~\cite[Sec. 2.1]{wisemanThoughtful2022}, featuring three observers: Alice, Bob, and, Charlie\footnote{The LF no-go theorem makes no assumption about what constitutes an \enquote{observer}. This specification is to be supplied by the user of the theorem or by a candidate theory.}. Bob and Charlie each measure their share of a bipartite system, obtaining outcomes labeled\footnote{We represent a variable by an upper case letter, and a specific value of the variable by the corresponding lower case letter.} $B$ and $C$, respectively. Charlie, who is in an isolated environment that we refer to as ``Charlie's laboratory'', performs a fixed measurement. Alice and Bob have choices of measurement settings labeled by $X$ and $Y$, respectively. Alice, who is stationed outside Charlie's laboratory, has the following choices: for $x=1$ she asks Charlie for his outcome and sets her own outcome to the value she heard from Charlie; for $x\neq 1$ she performs a different measurement on Charlie's laboratory, possibly disturbing or erasing the records of Charlie's result in the process.

Note that we have not demanded that Alice's and Bob's measurements are space-like separated here, since this is not needed for most of our results.
However, if they are space-like separated, so that the spatio-temporal relationships between the events $A$, $B$, $C$, $X$, and $Y$ are as given in Fig.~\ref{fig_spacetime}, we will call it the \textit{spacelike-separated} minimal LF scenario. In this case, just as in Bell's theorem, we have the background metaphysical assumption~\cite{wisemanCausarum2017} that such events are well-localized in space-time. In particular, $B$, $C$, and $Y$ must be outside the future light cone of $X$, while $A$, $C$, and $X$ must be outside the future light cone of $Y$. This is analogous to the locality-loophole-free Bell experiment. 

In~\cite{bongStrong2020,cavalcantiImplications2021,wisemanThoughtful2022}, it was shown that certain conjunctions of metaphysical assumptions on the spacelike-separated minimal LF scenario imply constraints on the operational correlations on $A$, $B$, $X$, and $Y$, known as the Local Friendliness (LF) inequalities. As detailed in \cite{cavalcantiImplications2021}, different sets of assumptions can be used to derive the LF inequalities, and we use the term \enquote{Local Friendliness} to refer to any such set of assumptions. One such pair of assumptions are \AOE\ and \LA.

\begin{definition}[\AOE ~(AOE)]
    Any observed event is an absolute single event, not relative to anything or anyone. In particular, there exists a joint probability distribution over all observed events.
\end{definition}

According to \AOE, all variables representing observed events (referred to as \emph{observed variables}) have well-defined single values in each run of the experiment. In the minimal LF scenario, this means, in particular, that there is an absolute event corresponding to Charlie's observed outcome $C$, even if Alice later erases it, and moreover that we can assign a well-defined joint probability distribution  $P(abc|xy)$ consistent with the marginals $P(ab|xy)=\sum_c P(abc|xy)$.

\begin{definition}[\LA ~(LA)]
\label{def_LA}
    If a measurement setting is freely chosen, then it is uncorrelated with any set of relevant events not in its future-light-cone.
\end{definition}

By ``freely chosen'' we mean that the measurement settings are independent variables that one has good background reasons to think are suitable to be used as inputs for randomized experimental trials. For example, the output of a random number generator can be considered a suitable ``free variable'' for this purpose. In the minimal LF scenario, we have Alice and Bob freely choose their measurement settings.  

\LA\ is motivated by relativity theory. In the spacelike-separated minimal LF scenario, under the assumption of \AOE, it demands that 
\begin{equation}
    \label{eq_LA}
        P(ac|xy)=P(ac|x), \quad P(bc|xy)=P(bc|y).
\end{equation}
Together with the assumption\footnote{This is naturally motivated by the protocol of the minimal LF scenario. It is a complementary assumption which has been called ``Tracking'' in \cite{schmid2023review}.} that ${P(a|c,x{=}1)=\delta_{a,c}}$ (since Alice simply asks Charlie for his outcome when $x=1$), one arrives at the LF inequalities.  The proof of this claim follows straightforwardly by noting that Eqs.~\eqref{eq_LA} imply all equations used in the derivation of LF inequalities in Ref.~\cite{wisemanThoughtful2022}. In this way, just like Bell inequalities, LF inequalities are derived in a theory-independent manner: that is, without assuming the validity of any particular theory (such as quantum theory). 

In~\cite{bongStrong2020,wisemanThoughtful2022}, quantum realizations of the minimal LF scenario are proposed where Alice is a superobserver who can perform (close enough to) arbitrary quantum operations, including operations that unitarily reverse\footnote{For such quantum realizations involving measurement reversing (which includes the erasure of Charlie's memory), one can imagine for concreteness that Charlie (called QUALL-E in \cite{wisemanThoughtful2022}) is an advanced AI running in a large future quantum computer. In this case \enquote{Charlie's laboratory} would involve all the qubits involved in the quantum computation instantiating QUALL-E.} Charlie's measurement when $x\neq 1$. Quantum theory predicts such a realization to violate LF inequalities.\footnote{Prototype experiments were performed in Refs.~\cite{proietti2019experimental,bongStrong2020} where (in hindsight for the case of \cite{proietti2019experimental}) LF inequalities were violated. For discussion on future experimental realizations involving increasingly complex and large systems, see~\cite{wisemanThoughtful2022}.} Thus the no-go result: if this quantum prediction holds, at least one of the metaphysical assumptions comprising Local Friendliness has to be false. 

\subsection{Overview of the results}

In this work, we first (in Sec.~\ref{sec_mr}) show that the LF inequalities are essentially special cases of monogamy relations, which are consequences of statistical marginal problems. That is, given that the joint distributions $P(abc|xy)$ (stipulated to exist by \AOE) need to satisfy the statistical constraints of Eqs.~\eqref{eq_LA} (stipulated by \LA), there must be a trade-off between the strength of the marginal correlation $P(ab|xy)$ and that of $P(ac|x{=}1)$ (where the latter is assumed to be a perfect correlation when deriving LF inequalities). 

\begin{figure}[htbp]
    \centering
   \includegraphics[width=0.15\textwidth]{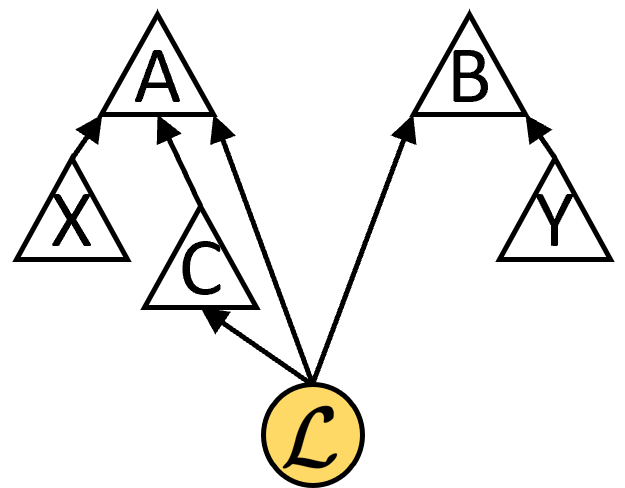}
    \caption{\textbf{The LF DAG.} The latent node $\cal{L}$ is highlighted in yellow as a reminder that our results hold even when the latent node is associated with nonclassical systems in the $d$-sep causal modeling framework.}
    \label{fig_LF}
\end{figure}

We then proceed to our main results concerning a framework for nonclassical causal inference that we term \emph{$d$-sep} causal modeling, which enables GPTs and certain post-GPTs to provide (a kind of) causal explanations. First, we define this framework in Sec.~\ref{sec_preliminaries}, while implicitly assuming Absoluteness of Observed Events by associating classical random variables to observed nodes. Then, in Sec.~\ref{sec_LFcm}, we show that the statistical marginal problem that leads to the LF inequalities can also be \emph{exactly} cast as a nonclassical causal marginal problem~\cite{greseleCausal2022,triantafillouLearning2010} phrased in the $d$-sep causal modeling framework. In this causal marginal problem, the LF inequalities are causal compatibility inequalities of the causal structure of Fig.~\ref{fig_LF}. This causal structure is the natural candidate for the causal structure given the experimental setup in the minimal LF scenario and is thus called the \emph{LF DAG}. 

Then, Sec.~\ref{sec_crisis} shows that---analogous to how Bell's theorem constitutes a serious challenge for classical causal inference~\cite{woodLesson2015}---the LF no-go theorem further challenges even \emph{nonclassical} causal inference. Firstly, we prove in Sec.~\ref{sec_RCA} that there is no way to explain the violation of LF inequalities with $d$-sep causal modeling if one wishes to retain two causal-metaphysical principles, i.e., \RCA\ and \IS, that underlie Local Agency~\cite{cavalcantiImplications2021}. Taken together with the result in Sec.~\ref{sec_LFcm}, this means that the LF DAG with $d$-sep causal modeling captures the essence of the LF theorem.  

Secondly, we prove in Sec.~\ref{sec_nft} that there is also no way to explain the violation of LF inequalities with $d$-sep causal modeling if one wishes to adhere to the No Fine-Tuning principle. Importantly, in Sec.~\ref{sec_cyc}, we further show that our No Fine-Tuning no-go theorem holds even when our $d$-sep causal modeling framework is extended to include cyclic causal structures. 

It is worth mentioning that, in the No Fine-Tuning argument of Sec.~\ref{sec_nft}, one of the conditional independence relations used in the No Fine-Tuning no-go theorem involves Charlie's measurement outcome even though it may be reversed by Alice afterwards. Nevertheless, we construct explicit protocols for gathering observational evidence for this statistical independence, which can be used in the previously mentioned quantum proposal for realizing violations of LF inequalities.

\section{LF inequalities are monogamy relations}
\label{sec_mr}

A \emph{statistical marginal problem}~\cite{vorobev_consistent_1962} concerns a set $\bm{P}$ of probability distributions over overlapping sets of variables. It studies whether there is a single joint probability distribution $P$ over the union of these sets that has the probabilities in $\bm{P}$ as its marginals, possibly under certain \emph{statistical} constraints. As we will see shortly, the original formulation~\cite{bongStrong2020} of the LF no-go theorem can be cast as a statistical marginal problem, in particular, one that yields monogamy relations.

\emph{Monogamy relations} are consequences of a special type of statistical marginal problem. In such problems, the joint distribution is the correlation among multiple parties in a given scenario, and the marginal probability distributions are the correlations shared between different subsets of parties. The monogamy relations are constraints one can derive on these marginal correlations under certain statistical conditions. They describe tradeoff relations between these marginals, wherein the correlation between one subset of the parties, say Alice and Bob, is in some way bounded by the correlation between another subset of the parties, say Alice and Charlie.

For example, in a tripartite Bell experiment, with the implicit assumption of \AOE~\cite{wisemanCausarum2017}, there is a joint distribution over all setting and outcome variables. Then, one can ask how strong the correlation between Alice and Bob can be, given certain degrees of correlation between Alice and Charlie, under the no-signaling condition. Here, the no-signaling condition is a statistical condition demanding that any of the marginal correlations between two parties be independent of the measurement setting of the third party. The consequences of such a statistical marginal problem can be expressed as inequalities on the marginal correlation between Alice and Bob together with that between Alice and Charlie; such inequalities are monogamy relations for the tripartite Bell scenario.

In our Local Friendliness scenario, as mentioned in Sec.~\ref{sec_intro}, the Absoluteness of Observed Events assumption demands there to be a well-defined joint distribution between the variables $A$, $B$, $C$, $X$, and $Y$. We now investigate how strong the correlation between Alice and Bob (namely, $P(ab|xy)$) can be, given the correlation between Alice and Charlie when $x=1$ (namely, $P(ac|x{=}1)$), under the statistical conditions in Eqs.~\eqref{eq_LA} that are implied by Local Agency. 

For example, in the case where all outcomes and settings are binary in the minimal LF scenario, the statistical constraints of Eqs.~\eqref{eq_LA} give rise to the following monogamy relation (as can be inferred from Ref.~\cite{augusiak2014elemental}):
\begin{equation}
    \label{CHSH_monogamy}
    {\rm CHSH} +\frac{1}{2} \sum_{a=c}P(ac|x{=}1) \leq \frac{5}{4},
\end{equation}
where 
\begin{align*}
    {\rm CHSH}=\frac{1}{4}\bigl[&\sum_{a=b} P(ab|00)+\sum_{a=b} P(ab|01) \\
    &+\sum_{a=b} P(ab|10)+ \sum_{a\neq b} P(ab|11)\bigr].
\end{align*}

Eq.~\eqref{CHSH_monogamy} illustrates a tradeoff relation between $P(ab|xy)$ and $P(ac|x{=}1)$: the larger the correlation between Alice's and Charlie's outcomes when $x=1$, the smaller the violation of the CHSH inequality that Alice and Bob can produce. In the special case where $a=c$ whenever $x=1$, Alice and Bob simply cannot violate the CHSH inequality. Although Eq.~\eqref{CHSH_monogamy} was originally obtained for the tripartite Bell scenario~\cite{augusiak2014elemental}, it applies to the minimal LF scenario because Eqs.~\eqref{eq_LA} (that come from Local Agency) are mathematically the same as the no-signaling conditions in the tripartite Bell scenario where Charlie does not have a choice of measurement setting.\footnote{Note that Eqs.~\eqref{eq_LA} cannot be simply viewed as the no-signaling conditions in the minimal LF scenario since Eqs.~\eqref{eq_LA} are not publicly available when records of Charlie's outcome $C$ are erased. Instead, in this case, the no-signaling conditions are just that  $P(a|xy)=P(a|x)$, and $P(b|xy)=P(b|y)$ (which are implied by \LA\ as \LA\ is a strictly stronger assumption). }

This example with the CHSH inequality arising from the minimal LF scenario with binary variables makes explicit what we mean when we say that the constraints that arise in the LF scenario are monogamy relations.\footnote{Another example is given by the so-called \enquote{relaxed} LF inequalities derived in Ref.~\cite[Eq.~(13)]{moreno2022events}. Mathematically, they are special cases of the monogamy relations for the Bell scenario derived in~\cite[Eq.~(3)]{augusiak2014elemental}.} In more generality, we can define the \emph{LF monogamy relations}:

\begin{definition}[LF monogamy relations]
\label{def_statMR}
    \emph{LF monogamy relations} are constraints on the set of probability distributions $\{P(ab|xy), P(ac|x{=}1)\}$ that arise from the requirement that these distributions are marginals of $P(abc|xy)$ satisfying the Local Agency constraints of Eqs.~\eqref{eq_LA}.
\end{definition}

As mentioned, the Local Friendliness inequalities are derived using the assumption that $P(a|c,x{=}1)=\delta_{a,c}$, since Alice simply reports Charlie's outcome when $x=1$. With this assumption, Eq.~\eqref{CHSH_monogamy} reduces to the CHSH inequality. Indeed, in the binary case, it was noted in Ref.~\cite{bongStrong2020} that the LF inequality is exactly the CHSH inequality.

Thus, the LF inequalities are special cases of LF monogamy relations; they are, at their core, consequences of a statistical marginal problem that involves the conditions of Eqs.~\eqref{eq_LA}.
\begin{definition}[LF inequalities]
\label{def_staLF}
    An \emph{LF inequality} is a special case of an LF monogamy relation, where ${P(ac|x{=}1)}$ is set to be $P(ac|x{=}1)=\delta_{a,c}P(c|x{=}1)$.
\end{definition}

In this work, we derive results for all LF monogamy relations instead of only the LF inequalities. This is a strengthening of the original LF no-go theorem, because it allows for the possibility of discrepancy between Charlie's actual observation and Alice's copy of his outcome (be it because of noise or other reasons~\cite{adlam2023does}). More explicitly, we can have 
\begin{equation}
\label{eq_techas}
\begin{split}
    \sum_{a\neq c}P(ac|x{=}1) := \gamma \geq 0
\end{split}\end{equation}
for some $\gamma\in[0,1]$. It may be hard to directly observe such a discrepancy, but we can instead qualify our conclusion for the statistics obtained in an experimental realization of the minimal LF scenario. That is, instead of immediately concluding that the LF assumptions are falsified whenever the LF inequalities are violated, we appeal to the LF monogamy relations to derive a lower bound on the discrepancy $\gamma$ which would be required to salvage the LF assumptions in light of the specific observed correlations $P(ab|xy)$. The option of holding onto the LF assumptions, then, gets weighed against the (im)plausibility of the lower bound on $\gamma$ computed from the monogamy relations.

Finally, bear in mind that in the definition of LF monogamy relations (Def.~\ref{def_statMR}) there is no assumption made regarding the type of system the three agents can share, such as classical, quantum, or another GPT.  This hints at the challenges our findings will present in terms of nonclassical causal inference being incapable of explaining violations of LF monogamy relations.

\section{LF no-go theorem as a nonclassical causal marginal problem}

\subsection{Nonclassical causal modeling}
\label{sec_preliminaries}

The main goal of causal inference is to find underlying causal mechanisms that explain the statistics of experiments or observations. The connection between the causal mechanisms and the observed statistics is given by a \emph{causal model}.

The class of causal models we consider here contains at least three parts. 

The first part is a \emph{causal structure}, represented by a graph $\mathcal G$. For simplicity, we assume for now that we are dealing with directed acyclic graphs (DAGs), although in Sec.~\ref{sec_cyc}, we will extend our framework to include cyclic causal models. We consider DAGs with two types of nodes, i.e., the \emph{observed nodes} (denoted by triangles) and the \emph{latent nodes} (denoted by circles); see for example Fig.~\ref{fig_LF}. An observed node is associated with an observed event, and is always associated with a classical random variable. In this sense, \AOE\ is embedded in this class of causal models. 

The second part is a \emph{theory prescription}, which is a specification of the type of theory under which probability distributions can be produced by the causal structure. Such a theory determines the types of system, e.g., classical, quantum, or GPT, that a latent node in $\mathcal G$ can be associated with, which in turn determines the compatibility criterion between $\mathcal G$ and a probability distribution. 

The third part is a probability distribution $P$ over the observed variables that is compatible with $\mathcal G$ given the theory prescription. We will expand on the topic of compatibility soon.

A causal model can also contain a specification of the quantitative causal relationship between the nodes in $\mathcal G$, represented by, for example, functions or channels. It can also include specific requirements for the systems associated with a latent node, such as its cardinality or Hilbert space dimension. Because these properties are not relevant for the results of this paper, we are not including them in the description here. This also makes it clearer that our results are device-independent in the sense that they do not rely on any detailed assumptions about systems, channels, or instruments. 

A causal model with a causal structure $\cal G$ is called a \emph{classical causal model} when its theory prescription is classical theory. Under such a rule, all nodes in $\cal G$ must be associated with classical random variables. A probability distribution $P$ that can be produced by $\cal G$ under classical theory is said to be \emph{classical-compatible} with $\mathcal G$. The classical-compatibility criterion is often expressed as the classical Markov condition~\cite{Pearl_2021}. We use the term \emph{GPT causal model} to refer to any causal model whose theory prescription is some GPT and consequently, whose latent nodes can be associated with those GPT systems. Appendix~\ref{app_GPTcom} provides details on how a probability distribution can be produced by a causal structure given the GPT, following Ref.~\cite{hensonTheoryindependent2014}. As long as there is \emph{some} GPT that allows a probability distribution $P$ to be produced by $\cal G$, we say that $P$ is \emph{GPT-compatible} with $\cal G$. When the GPT is specified to be quantum theory, such a GPT causal model is a \emph{quantum causal model}, where any latent node in $\cal G$ is allowed to be associated with a quantum system.  

It is also possible to have compatibility rules with even more general nonclassical probabilistic theories and to define a wider class of causal models that we call the \emph{$d$-sep causal models}. However, to spell out what we mean by $d$-sep causal models, it is instructive to first see some common features of the compatibility criteria specified by all GPTs.

In general, the constraints that a causal structure imposes on its compatible probability distributions depend on the theory prescription. For example, the Bell inequality is a constraint of the Bell DAG under classical causal modeling but is not a constraint if the theory prescription is quantum or an arbitrary GPT. However, some constraints can hold across various theory prescriptions, of which the most well-known are conditional independence relations between observed variables implied by $d$-separation relations~\cite{Geiger1988,verma_pearl}, which are graph-theoretic relations of the causal structure. For example, the no-signaling conditions in a Bell scenario are constraints of this form, and so hold for the Bell DAG in all GPT causal models. (See Appendix~\ref{appendix_dsep} or \cite[Chapter 1]{causality_pearl} for a definition and discussion of $d$-separation.) 

A causal structure $\cal G$ together with a theory prescription is said to obey the \emph{$d$-separation rule}\footnote{Also called the \enquote{directed global Markov property}, for example, in Ref.~\cite{forre2017markov}.} when any $d$-separation relation among observed nodes of the causal structure implies a conditional independence relation on its compatible probability distributions. That is, for any three sets of observed nodes, $\bm{U}$, $\bm{V}$, and $\bm W$ in $\cal G$, whenever $\bm{U}$ and $\bm{V}$ are $d$-separated by $\bm W$ (denoted as ${\bm{U}\perp \bm{V}|\bm{W}}$), then the conditional independence relation $P{(\bm{u}|\bm{v},\bm{w})=P(\bm{u}|\bm{w})}$ must hold for all distributions compatible with $\mathcal G$. When the graph is acyclic, i.e, a DAG, it was proven in Ref.~\cite{hensonTheoryindependent2014} that the $d$-separation rule is satisfied regardless of the GPT specified by the theory prescription.

In light of this, we say that a probability distribution $P$ is \emph{$d$-sep-compatible} with a DAG $\cal G$ if $P$ exhibits all of the conditional independence relations coming from the $d$-separation relations among observed nodes of $\cal G$, and we define:
\begin{definition}[$d$-sep causal model]
    Any causal model whose probability distribution and causal structure are $d$-sep-compatible is a \emph{$d$-sep causal model}.
\end{definition}
It has been shown that GPT causal models cannot always explain every distribution which is merely $d$-sep-compatible with a causal structure~\cite{hensonTheoryindependent2014}; that is, for certain causal structures, there exists probability distributions that are $d$-sep-compatible but are nevertheless not GPT-compatible. We have been made aware that quasiprobabilistic causal models---wherein the real number assigning the (quasi)probability of a particular latent valuation need not lie between zero and one\footnote{A quasiprobabilistic model is slightly more general than a GPT causal model, since in a GPT every possible complete circuit one can consider must be a valid probability, while in a quasiprobabilistic model one need only demand that the specific circuits in the model generate valid probabilities over the observed nodes.}---on a DAG $\cal G$ can give rise to distributions that are not GPT-compatible with $\cal G$.\footnote{The example provided to us is that, with quasiprobabilistic theories, it is possible to achieve perfect correlation between the three variables of the so-called triangle scenario~\cite{privateMarc}. Such perfect correlation cannot be realized by GPTs in that scenario.} Therefore, there exist theory prescriptions that go beyond GPTs and are nevertheless included in our $d$-sep causal modeling framework. 
For reference, $d$-sep causal models coincide with the so-called Ordinary Markov models~\cite{shpitserIntroduction2014}, and the set of probability distributions $d$-sep-compatible with an acyclic causal structure is the set ${\cal I}$ defined in Ref.~\cite{hensonTheoryindependent2014}.

We postpone to Sec.~\ref{sec_cyc} the introduction of an extension of $d$-sep causal models,  encompassing a wide class of cyclic causal models, namely, those adhering to a generalization of the $d$-separation rule. Importantly, all our findings, including the ones prior to Sec.~\ref{sec_nft} remain valid in that extended framework.

 We emphasize again that the assumption of \AOE\ is embedded in all the causal modeling frameworks we consider in this paper. 

\emph{Causal marginal problems}---a special case of causal compatibility problems---have been studied by the classical causal inference community for more than two decades~\cite{Danks2002,triantafillouLearning2010}. Here, we define these in such a way that they also apply to nonclassical causal inference. Similar to a statistical marginal problem, a causal marginal problem concerns a set $\bm{P}$ of probability distributions over non-identical but overlapping sets of \emph{observed} variables. It concerns whether there is a single joint probability distribution $P$ over the union of these sets of variables that has the distributions in $\bm{P}$ as its marginals under certain \emph{causal} constraints. Such causal constraints can be, for example, that $P$ must be compatible with a causal structure $\cal{G}$ under a certain theory prescription. When there exists such a $P$ (that reproduces $\bm{P}$) that is compatible with $\cal{G}$ under that theory prescription, we say that the \emph{set} $\bm{P}$ itself is compatible with $\cal{G}$, meaning that all probability distributions in the set are \emph{jointly} compatible with $\cal{G}$.

\subsection{LF inequalities are nonclassical causal compatibility inequalities}
\label{sec_LFcm}

Now, we demonstrate how and why the statistical marginal problem of the LF no-go theorem, as introduced in Sec.~\ref{sec_mr}, can be \emph{exactly} cast as a causal marginal problem relative to the LF DAG, meaning that the formulation of the no-go theorem using the LF DAG does not lose or gain any constraints on operational data than what was already in the original formulation of the LF no-go theorem \cite{bongStrong2020}.

First, we prove that LF inequalities are causal compatibility inequalities under $d$-sep causal modeling.
\begin{theorem}[No-go---LF DAG]
\label{thm_LF_DAG}
    No $d$-sep causal model with the LF DAG (Fig.~\ref{fig_LF}) can explain any violation of LF monogamy relations, including LF inequalities. 
\end{theorem}
\begin{proof}
The LF DAG has the following $d$-separation relations:
    \begin{equation}
        \label{eq_dLA}
        AC\perp Y|X, \quad   BC\perp X|Y,
    \end{equation}
    By the $d$-separation rule, Eq.~\eqref{eq_dLA} demands the conditional independence relations of Eqs.~\eqref{eq_LA} to hold for any distribution $d$-sep-compatible with $\mathcal{G}$. But such a distribution must satisfy the LF monogamy relations, by Def.~\ref{def_statMR}.
\end{proof}

Theorem~\ref{thm_LF_DAG} is a formulation of the LF no-go theorem from the perspective of a causal marginal problem, which is a special case of a causal compatibility problem. Furthermore, the following theorem implies that such a perspective with the LF DAG does not impose any extra constraints on $\{P(ab|xy), P(ac|x{=}1)\}$ beyond those from the statistical marginal problem in Def.~\ref{def_statMR}. This theorem is stated in terms of GPT causal models, thus showing that there are no extra constraints even when we consider only GPT-compatibility (instead of the more general $d$-sep-compatibility).
\begin{theorem}
\label{thm_only}
    The \emph{only} constraints on the sets of probability distributions $\{P(ab|xy), P(ac|x{=}1)\}$ that are GPT-compatible with the LF DAG are the LF monogamy relations. 
\end{theorem}

\begin{proof}

\begin{figure}[h]
    \centering	\includegraphics[width=0.15\textwidth]{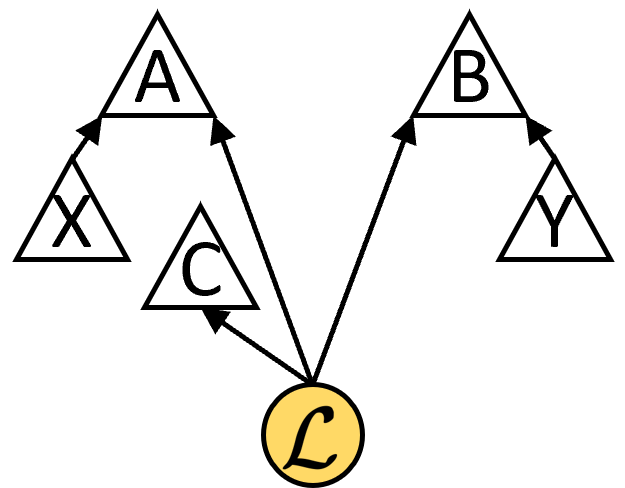}
	\caption{The DAG for a tripartite Bell experiment where one of the three observers, Charlie, does not have choices for his measurement setting.} 
	\label{fig_triDAG}
	\end{figure}
 
Consider the DAG in Fig.~\ref{fig_triDAG}, which represents a tripartite Bell experiment where one of the three observers, Charlie, does not have choices for his measurement setting. Ref.~\cite{barrett2006information} has an explicit construction of a GPT, called the Generalized Non-signaling Theory, such that when the latent node $\cal{L}$ is associated with it, the DAG in Fig.~\ref{fig_triDAG} is compatible with \emph{any} $P(abc|xy)$ satisfying Eqs.~\eqref{eq_LA}. 

The LF DAG, compared to the DAG in Fig.~\ref{fig_triDAG}, has an additional arrow from $C$ to $A$. Thus, the LF DAG must have no less power than 
the DAG in Fig.~\ref{fig_triDAG} in explaining the correlations of $P(abc|xy)$, under any $d$-sep theory prescription.\footnote{In fact, one can prove the LF DAG and the DAG in Fig.~\ref{fig_triDAG} are equivalent under any $d$-sep theory prescription in explaining $P(abc|xy)$. } Thus, it must be GPT-compatible with all the probability distributions that are GPT-compatible with the DAG of Fig.~\ref{fig_triDAG}. Thus, the LF DAG must be GPT-compatible with \emph{any} $P(abc|xy)$ satisfying Eqs.~\eqref{eq_LA}. Consequently, by Def.~\ref{def_statMR}, the LF DAG must be GPT-compatible with \emph{any} set of $\{P(ab|xy), P(ac|x{=}1)\}$ satisfying the LF monogamy relations.
\end{proof}

Note that by opting to present Theorem~\ref{thm_LF_DAG} in terms of $d$-sep causal modeling and Theorem~\ref{thm_only} in terms of GPT causal modeling, we are presenting the most general version of both theorems: Theorem~\ref{thm_LF_DAG} says that even the most general $d$-sep causal model one could think of would not help to explain the violation of the LF inequalities; meanwhile, Theorem~\ref{thm_only} says that even by restricting the $d$-sep causal models to only GPT ones, there are still no extra constraints imposed on $\{P(ab|xy), P(ac|x{=}1)\}$ by the LF-DAG other than LF monogamy relations. Together, these theorems imply that there is no difference between the set of probability distributions that are GPT-compatible and those that are $d$-sep-compatible with the LF DAG.

Therefore, we do not gain or lose any constraint on the operational probability $P(ab|xy)$ in the minimal LF scenario by casting the LF no-go theorem in terms of a causal marginal problem with the LF DAG. This is true whether the causal marginal problem is framed within the framework of GPT causal modeling or that of $d$-sep causal modeling.

\section{The conundrum for nonclassical causal inference}
\label{sec_crisis}

\begin{figure}[!h]
\captionsetup[subfigure]{aboveskip=-2pt,belowskip=-1pt}
    \centering
    \begin{subfigure}[b]{0.13\textwidth}
         \centering
         \includegraphics[width=\textwidth]{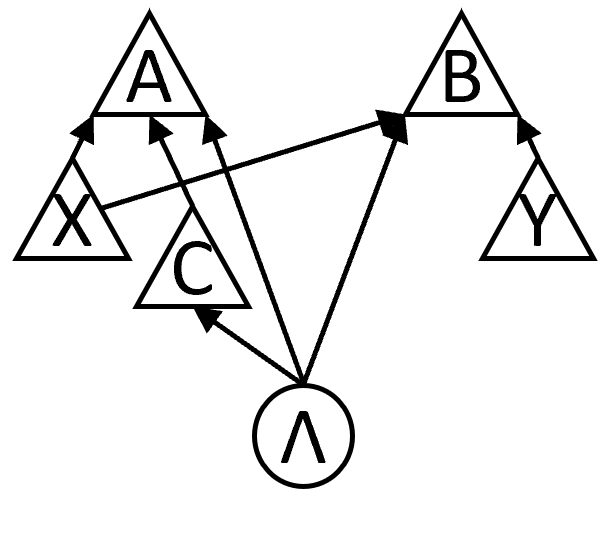}
         \caption{}
    \label{fig_LF-noloc}
     \end{subfigure}
    \hspace{1mm}
    \begin{subfigure}[b]{0.13\textwidth}
         \centering
         \includegraphics[width=\textwidth]{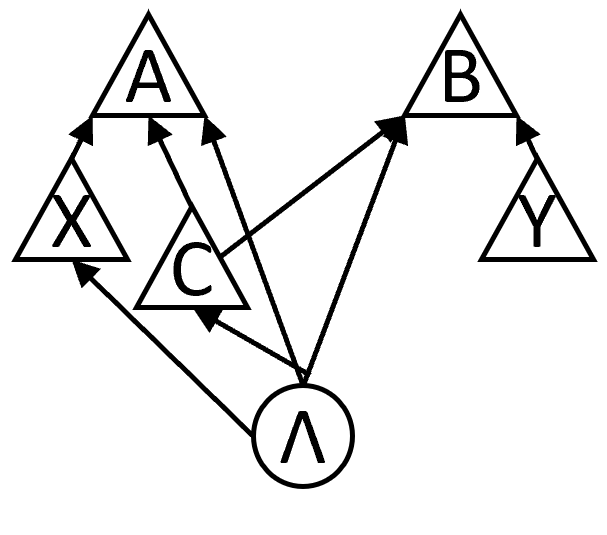}
         \caption{}
         \label{fig_LF-SD}
     \end{subfigure}
     \hspace{1mm}
    \begin{subfigure}[b]{0.13\textwidth}
         \centering
         \includegraphics[width=\textwidth]{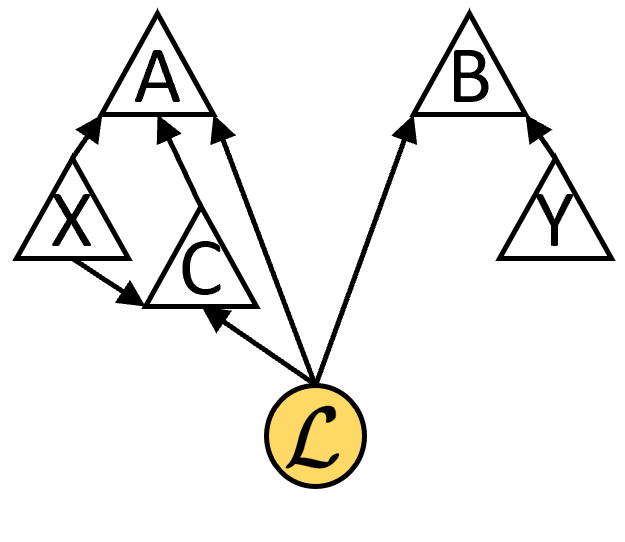}
         \caption{}
         \label{fig_LF-RetroCau}
     \end{subfigure}
    \caption{\textbf{Examples of causal structures allowing for violations of LF inequalities}.}
    \label{fig_LF-sol}
\end{figure}

Even though the LF DAG cannot explain any violation of LF monogamy relations under $d$-sep causal modeling, one could still hope to provide a causal explanation by changing the causal structure (i.e., the DAG). Fig.~\ref{fig_LF-sol} shows some examples of potential alternative causal structures. For the causal structures in Figs.~\ref{fig_LF-noloc} and~\ref{fig_LF-SD}, there are even \emph{classical} causal models that can explain the violation of LF monogamy relations; for Fig.~\ref{fig_LF-RetroCau}, it is necessary to use nonclassical theory prescriptions.

In the case of Bell's theorem, Ref.~\cite{woodLesson2015} showed that, besides being in tension with relativity, all \emph{classical} causal models that can explain the violation of Bell inequalities are fine-tuned. In this section, we prove an analogous result for any \emph{$d$-sep} causal model that can explain the violation of LF monogamy relations: all of them need to simultaneously violate causal-metaphysical assumptions motivated by, for example, relativity, \emph{and} be fine-tuned. It is then clear why the problem for causality posed by the LF theorem is much stronger than the one posed by Bell's theorem: while the latter is only an issue for classical causal inference, the LF theorem calls even the much more general existing frameworks for {\em nonclassical} causal inference in question.

\subsection{In tension with relativity}
\label{sec_RCA}

The theorem that will be presented in this section concerns the \emph{spacelike-separated} minimal LF scenario, as it uses a causal-metaphysical assumption motivated by relativity. These causal-metaphysical assumptions are causal motivations for \LA\ (see Def.~\ref{def_LA}). They were previously defined in Ref.~\cite{cavalcantiImplications2021} as
\begin{definition}[\RCA]
		Any cause of an event is in its past light cone.
\end{definition}
\begin{definition}[\IS]
		A setting has no relevant causes i.e., it can always be chosen via suitable variables that do not have causes among, nor share a common cause with, any of the other experimental variables.
\end{definition}
\RCA\ is motivated by the theory of relativity, while \IS\footnote{This assumption was called \emph{Independent Interventions} in~\cite{cavalcantiImplications2021}. However, since \enquote{intervention} has a specific technical meaning in causal inference literature, we will use the term \IS\ for the scope of this work.}
is motivated by the common assumption that Alice and Bob can freely choose their measurement settings. 

\begin{theorem}[No-go---relativistic]
\label{thm_RCA}
    No $d$-sep causal model satisfying \RCA\ and \IS\ can explain any violation of the LF monogamy relations (including the LF inequalities) in the spacelike-separated minimal LF scenario.
\end{theorem}
\begin{proof}
As introduced in Sec.~\ref{sec_intro}, in the spacelike-separated minimal LF scenario, $B$, $C$, and $Y$ must be outside the future light cone of setting $X$, and $A$, $C$, and $X$ must be outside the future light cone of setting $Y$. 

\IS\ demands that the settings $X$ and $Y$ have no relevant causes. Specifically, in a DAG that obeys \IS, $X$ and $Y$ cannot be caused by nor share common causes with any other node. Thus, they can only be not-$d$-separated from a node if they are causes of that node.

\RCA\ demands that $X$ cannot be a cause of $B$, $C$, and $Y$, while $Y$ cannot be a cause of $A$, $C$, and $X$. This then implies that any causal structure that obeys both \RCA\ and \IS\ needs to exhibit the following $d$-separation relations:
\begin{equation}
    B\perp X, \quad C\perp X, \quad A\perp Y, \quad C\perp Y, \quad X\perp Y.
\end{equation}
By the compositionality property of $d$-separation (see Def.~\ref{def_comp} for the definition of compositionality, and Appendix~\ref{appendix_dsep} for a proof that $d$-separation is compositional), these imply that
\begin{equation}
    ACX\perp Y, \quad BCY\perp X.
\end{equation}
The $d$-separation rule demands 
\begin{equation}
    P(acx|y)=P(y), \quad P(bcy|x)=P(bcy).
\end{equation}
These equations imply Eqs.~\eqref{eq_LA}, which yields the LF monogamy relations by Def.~\ref{def_statMR}.
\end{proof}

Theorem~\ref{thm_RCA} reformulates the challenge raised in Ref.~\cite{cavalcantiImplications2021} to quantum causal models, while strengthening it to challenge all $d$-sep causal models. Furthermore, Theorem~\ref{thm_RCA} together with Theorem~\ref{thm_only} imply that

\begin{corollary}
    \label{corollary_relativistic}
The set of $P(abc|xy)$ that are GPT-compatible with the LF DAG is a superset of that of any $d$-sep causal model satisfying \RCA\ and \IS.
\end{corollary}

That is, the LF DAG under GPT causal modeling provides the most powerful $d$-sep causal explanation for the correlations of $P(abc|xy)$ while adhering to \RCA\ and \IS.\footnote{More precisely, the LF DAG is a representative of the equivalence class of DAGs under $d$-sep causal modeling that are the most powerful (i.e., it can explain any probability distribution that is $d$-sep-compatible with any DAG satisfying \RCA\ and \IS). This class includes, for example, the GPT causal model with the DAG of Fig.~\ref{fig_triDAG}. It is an equivalence class since all DAGs in it are $d$-sep-compatible with the same set of probability distributions.}

\subsection{At odds with No Fine-Tuning}
\label{sec_nft}

As mentioned, Theorem~\ref{thm_RCA} applies to observations made in the spacelike-separated minimal LF scenario. Therefore, when considering \emph{non}-spacelike-separated scenarios, Theorem \ref{thm_RCA} has no teeth. Nevertheless, there is an independent challenge to models that attempt to explain the violation of LF monogamy relations via an alternative causal structure: they violate the No Fine-Tuning condition (also known as the faithfulness condition), which is an assumption often used in causal discovery.

\begin{definition}[No Fine-Tuning / Faithfulness]
\label{def_nft}
    A causal model is said to satisfy \emph{No Fine-Tuning} or \emph{Faithfulness} if every conditional independence relation in its probability distribution comes from the corresponding $d$-separation relation in its causal structure.
\end{definition}

In a fine-tuned model, the causal structure is not sufficient to explain why some conditional independences exist, which calls for further explanations. Cryptographic protocols are examples of fine-tuned processes with a supplementary explanation; see Appendix~\ref{appendix_cypher} for an example.

In Refs.~\cite{woodLesson2015,cavalcantiClassical2018,Pearl_2021} it was proven that all classical causal models reproducing violations of Bell inequalities must be fine-tuned to satisfy the no-signaling conditions in a Bell scenario. Here, we construct a No Fine-Tuning no-go theorem for the minimal LF scenario, where not only classical causal models are considered, but also any nonclassical ones under $d$-sep causal modeling, with extensions to cyclic causal structures in Sec.~\ref{sec_cyc}.

If we assume that it is physically impossible to send signals faster than light, accepting fine-tuned models for a Bell scenario comes at the extra cost of being unable to single out a unique correct causal structure \emph{even in principle} (as we expand upon by contrasting with the cryptographic example in Appendix~\ref{appendix_cypher}, where one can single out a unique causal structure in principle). 
The results of Refs.~\cite{woodLesson2015,cavalcantiClassical2018,Pearl_2021} then imply that \emph{all} classical causal models for a Bell scenario, apart from requiring a supplementary explanation for no-signaling, also suffer from this extra burden. Here we show the same for $d$-sep causal models for the minimal LF scenario.

For our argument, besides the usual no-signaling conditions, namely,
\begin{equation}
           P(a|xy)=P(a|x),  \quad    P(b|xy)=P(b|y), \label{eq_signaling} 
\end{equation} 
we will also use an additional conditional independence relation that can be expected for the minimal LF scenario, namely, the statistical independence of Alice's or Bob's measurement setting from Charlie's outcome:
\begin{align}
    P(c|xy)=P(c).  \label{eq_CXY}
\end{align}

Both Eqs.~\eqref{eq_signaling} and \eqref{eq_CXY} can be motivated by observations. The data to verify Eqs.~\eqref{eq_signaling} is always publicly available at the end of an experimental realization of the minimal LF scenario. Violation of Eq.~\eqref{eq_CXY} would mean that upon observing $c$, Charlie can obtain information about Alice's and/or Bob's future choices of setting; as such, Eq.~\eqref{eq_CXY} is also a kind of no-signaling condition. However, this data may not always be publicly available in some realizations of the minimal LF scenario such as the quantum proposal mentioned in Sec.~\ref{sec_intro}, because in such realizations the record of Charlie's outcome might be erased by Alice. Nevertheless, observational evidence of Eq.~\eqref{eq_CXY} can still be publicly gathered in these realizations. Protocols for verifying Eq.~\eqref{eq_CXY} in such a quantum proposal will be presented in Sec.~\ref{sec_vero}, and some alternative protocols are presented in Appendix~\ref{app_nft}. 

It turns out that the requirement of No Fine-Tuning rules out any causal structure that is $d$-sep-compatible with violations of LF inequalities.

\begin{theorem}[no-go---No Fine-Tuning]
\label{thm_nft}
    No $d$-sep causal model that satisfies No Fine-Tuning, ${P(a|xy)=P(a|x)}$, ${P(b|xy)=P(b|y)}$, and ${P(c|xy)=P(c)}$ can explain any violation of the LF monogamy relations (including the LF inequalities). 
\end{theorem}

\begin{proof}
${P(c|xy)=P(c)}$ implies that ${P(c|xy)=P(c|x)}$ and ${P(c|xy)=P(c|y)}$. No Fine-Tuning requires the DAG reproducing $P(a|xy)=P(a|x)$, $P(b|xy)=P(b|y)$. ${P(c|xy)=P(c|x)}$ and ${P(c|xy)=P(c|y)}$ to satisfy the $d$-separation relations of     
\begin{gather}
	A \perp Y|X,  \, B \perp X|Y, \,  C \perp X|Y, \, C \perp Y|X. 
    \end{gather}
Because of the compositionality property of $d$-separation (see Def.~\ref{def_comp} for the definition of compositionality, and Appendix~\ref{appendix_dsep} for a proof that $d$-separation is compositional), the above $d$-separation relations imply that
\begin{equation}
        AC\perp Y|X, \quad   BC\perp X|Y.
\end{equation}
By the $d$-separation rule, these imply Eqs.~\eqref{eq_LA}, which yields the LF monogamy relations by Def.~\ref{def_statMR}.
\end{proof}

That is, as long as one would like their causal model for the minimal LF scenario to reproduce the conditional independence relations of Eqs.~\eqref{eq_signaling} and~\eqref{eq_CXY}, for whatever motivations one may have (including operational evidence or metaphysical considerations) Theorem~\ref{thm_nft} demonstrates that all $d$-sep models that reproduce violations of LF monogamy relations must be fine-tuned. Thus, Theorem~\ref{thm_nft} provides a novel way of understanding violations of LF inequalities---not only do they contradict relativistic causal-metaphysical assumptions in spacelike-separated minimal LF scenarios, but also challenge \emph{any} $d$-sep causal model that respects the causal discovery assumption of No Fine-Tuning.

Just as in the case in Sec.~\ref{sec_RCA}, we have the following corollary from Theorem~\ref{thm_only} and Theorem~\ref{thm_nft}.
\begin{corollary}
The set of $P(abc|xy)$ that are GPT-compatible with the LF DAG is a superset of that of any $d$-sep causal model satisfying Eqs.~\eqref{eq_signaling} and~\eqref{eq_CXY} and No Fine-Tuning.
\end{corollary}

That is, under No Fine-Tuning, we again arrive at the LF DAG---with GPT causal modeling, it already provides the most powerful $d$-sep causal model one can obtain for the distribution $P(abc|xy)$ while reproducing the desired conditional independences of Eqs.~\eqref{eq_signaling} and~\eqref{eq_CXY}.

\subsubsection{Applicability to cyclic causal models}
\label{sec_cyc}

Until this point, we only worked with acyclic causal structures, leaving open the question of whether causal models with cyclic causal structures can explain LF inequality violations without violating the causal-metaphysical assumptions of Theorem~\ref{thm_RCA} or No Fine-Tuning. However, it is clear that cyclic causal structures directly violate \RCA. Moreover, as we will prove in this subsection, Theorem~\ref{thm_nft} can be extended to a wide class of cyclic causal models. We do this by generalizing the framework of $d$-sep causal modeling to a framework defined here as \emph{compositional causal modeling}, which employs graphical-separation rules more general than the $d$-separation rule.  

Recall from Sec.~\ref{sec_preliminaries} that the $d$-separation rule means that $d$-separation relations among observed nodes imply conditional independence relations in any compatible probability distribution. However, this rule is not necessarily valid for a cyclic causal model\footnote{We note that there are cyclic causal structures that obey the $d$-separation rule with \textit{all} theory prescriptions. This class is non-trivial since it contains causal models that can violate causal inequalities detecting indefinite causal order~\cite{Oreshkov_2012}. One example is the Lugano process~\cite{Baumeler_2014}, which trivially satisfies the $d$-separation rule since it has no $d$-separation relation.}; see Appendix~\ref{appendix_cyclic} for an example.
Because of that, generalizations of the $d$-separation rule have been developed to encompass a larger number of cyclic causal structures~\cite{Geiger1988,forre2017markov,Carla2023}. We use the term \emph{graphical-separation relations} to denote the graph-theoretic relations employed in those generalizations. A causal structure together with a theory prescription is said to obey a \emph{graphical-separation rule} when any of the corresponding graphical-separation relations among observed nodes implies a conditional independence relation on its compatible probability distributions. 
 
Examples of graphical-separation rules that generalize the $d$-separation rule includes the $\sigma$-separation rule~\cite{forre2017markov} and the $p$-separation rule~\cite{Carla2023}.  While the $\sigma$-separation rule can still be violated by some cyclic causal structures with classical theory prescription (as shown in the example of Appendix~\ref{appendix_cyclic}), $p$-separation is proved to be valid at least for \emph{all} causal structures under \textit{quantum} theory prescription~\cite{Carla2023}. Both the $\sigma$-separation rule and the $p$-separation rule reduce to the $d$-separation rule for acyclic causal structures.

A common feature of $d$-separation, $\sigma$-separation, and $p$-separation rules is their compositionality, which is also a key ingredient needed for the proof of the No Fine-Tuning theorem (Theorem~\ref{thm_nft}).
\begin{definition}[Compositional graphical-separation relations]
\label{def_comp}
    A graphical-separation relation denoted by $\oldperp$ is compositional if $\boldsymbol{U}\oldperp \boldsymbol{W}|\boldsymbol{Z}$ and $\boldsymbol{V}\oldperp \boldsymbol{W}|\boldsymbol{Z}$ implies that $\boldsymbol{UV}\oldperp \boldsymbol{W}|\boldsymbol{Z}$, where $\boldsymbol{U}$, $\boldsymbol{V}$, $\boldsymbol{W}$, and $\boldsymbol{Z}$ are four sets of observed nodes in a graph.
\end{definition}
The fact that $d$-, $\sigma$-, and $p$-separation relations are compositional follows immediately from their definitions~\cite{forre2017markov,Carla2023}, where the set-wise relations are defined by the element-wise relations:  Given a set $\boldsymbol{Z}$, two sets $\boldsymbol{U}$ and 
 $\boldsymbol{W}$ satisfy ${\boldsymbol{U}\oldperp\boldsymbol{W}|\boldsymbol{Z}}$ if and only if ${\forall U_i\in\boldsymbol{U}}, {\forall W_j\in\boldsymbol{V}}$, ${U_i\oldperp W_j|\boldsymbol{Z}}$;
see Def.~\ref{def_dsep} in Appendix~\ref{appendix_dsep} for example. This suggests that compositionality is a natural property of graphical-separation relations. Furthermore, the breadth of existing examples suggests that every GPT cyclic causal model should satisfy some compositional separation rule. 

A compositional causal model is any causal model that adheres to a compositional graphical-separation rule (and where its observed nodes are associated with classical random variables). The \enquote{minimum definition of a causal model} in Ref.~\cite{Vilasini_2022} is a special case of a compositional causal model, wherein the graphical-separation rule is simply $d$-separation and cyclic causal models are permitted.

We can now extend the notion of No Fine-Tuning to compositional causal models. 
\begin{definition}[No Fine-Tuning / Faithfulness (compositional causal model)]
\label{def_nft2}
    Suppose that a causal model satisfies a graphical-separation rule with a compositional separation relation $\oldperp$. The model is said to satisfy \emph{No Fine-Tuning} or \emph{Faithfulness} if every conditional independence relation in its probability distribution corresponds to a $\oldperp$ relation in its causal structure.
\end{definition}
Clearly, this notion reduces to Def.~\ref{def_nft} if $\oldperp$ is $\perp$. With this updated definition of No Fine-Tuning, the proof of Theorem~\ref{thm_nft} is still valid when we replace $\perp$ by $\oldperp$. 
\begin{theorem}[no-go---No Fine-Tuning (compositional causal model)]
\label{thm_nftcomp}
    No compositional causal model that satisfies No Fine-Tuning,  ${P(a|xy)=P(a|x)}$, ${P(b|xy)=P(b|y)}$, and ${P(c|xy)=P(c)}$ can explain any violation of the LF monogamy relations (including the LF inequalities). 
\end{theorem}

\subsubsection{Verifying the statistical independence of an inaccessible variable}
\label{sec_vero}

Now, we will describe two different verification protocols for checking the validity of Eq.~\eqref{eq_CXY}, i.e., ${P(c|xy)=P(c)}$, in the proposed quantum realization of violations of LF inequalities mentioned in Sec.~\ref{sec_LFscenario}, where Alice sometimes reverses Charlie's measurement. Note that the correctness of our No Fine-Tuning no-go results does not rely on these verification protocols, but the latter provide a motivation from an operational perspective for having $P(c|xy)=P(c)$ in the causal models for the minimal LF scenario even when this independence relation may not be publicly observable.

These verification protocols require the runs of quantum LF experiments to be executed simultaneously, i.e., in a parallel instead of sequential fashion, with multiple Charlies. They also involve another agent, Veronika\footnote{Besides the fact that \enquote{vero} means \enquote{true}, the name of the verifier is a tribute to Veronika Baumann, who pointed out that the verifier in our verification protocols can send out a verification result that needs not be erased afterward, which is inspired by Deutsch's protocol for Wigner's friend~\cite{deutschQuantum1985} and her work~\cite{baumann2019wigners,baumann2023observers}.} the verifier. She receives the necessary data to verify that Eq.~\eqref{eq_CXY} is satisfied and communicates the verification result to the rest of the world, \emph{without} communicating any of the individual values of the Charlies' outcomes. 

In both protocols, assume that we execute $N$ parallel runs of the experimental realizations for the minimal LF scenario. At the beginning of the protocol, $N$ identical bipartite systems are prepared. Instead of one Charlie, we have $N$ Charlies, each receiving half of a bipartite system. The remaining halves are all sent to Bob. Each Charlie is in a lab isolated from the others and measures his half in a fixed basis; they do not share their outcomes with each other. Then, Alice makes her measurement choices for all $N$ parallel runs, by using, e.g., a random number generator to generate $N$ values of $X$, each of which corresponds to her measurement setting in each of the $N$ runs; similarly, Bob makes his measurement choices for all $N$ runs of $Y$, each of which corresponds to his measurement setting in each run.

Before Alice implements any measurements, Veronika checks if $C$ is independent of $X$ and $Y$. She has two ways to do so. We will first present a quantum verification protocol in which Veronika is a superobserver with quantum control over all of the Charlies' labs. Then, we will present a theory-independent verification protocol where Veronika does not need to be a superobserver. In both protocols, Veronika collects all values of $X$ and $Y$ from Alice and Bob, but the way she interacts with the Charlies is different.

For both variants, we will show how the addition of the verification process will not prevent the violation of LF inequalities in a quantum realization. Our analysis involves idealizations made by appealing to the limit where $N\rightarrow\infty$. However, this is not of great concern since even for observing the no-signaling condition, one always needs certain idealizations to convince oneself that, even though the perfect independence relation such as $P(a|xy)=P(a|x)$ is never observed in a finite ensemble, the no-signaling condition is nevertheless verified in the lab. In Appendix~\ref{app_nft}, we provide our responses to some other concerns one may have regarding the analysis. 

We emphasize again that the correctness of our No Fine-Tuning no-go results does not rely on the feasibility or the validity of the verification protocols---the verification protocols proposed here merely serve as an operational motivation for having $P(c|xy)=P(c)$ in a nonclassical causal model for the minimal LF scenario.

\medskip
  
\textbf{A quantum verification protocol}

In the first protocol,  Veronika is assumed to be a superobserver who can perform quantum operations on all of the Charlies' labs; in fact, Alice can take the role of Veronika. After the Charlies make their measurements, and Alice and Bob make their measurement choices, Veronika implements a dichotomic projection-valued measurement (PVM) on the joint system consisting of all the Charlies' labs, without asking for individual values of $C$. In such a PVM, one of the projectors (coming in Eq.~\eqref{eq:projectPass}) corresponds to ${\forall c, x, y:|f(c|xy)-f(c)|<\epsilon}$, where $f(\cdot)$ indicates relative frequencies and $\epsilon$ is a small positive value indicating the allowed statistical fluctuations. If the measurement outcome corresponds to this projector, Veronika records the verification result as \enquote{pass}, and otherwise, as \enquote{fail}. Then, Alice proceeds with her measurement on each Charlie and his share of the bipartite system, and Bob proceeds with his measurement on each of Bob's share of the bipartite system.

This first protocol relies on the quantum description of the PVM, to which we now turn. Given the validity of quantum theory, when $N$ is large enough, the verification process performed by Veronika does not prevent the quantum violation of LF inequalities. Here we outline the proof of this statement.

Let the initial state of the bipartite system shared between Charlie$_i$ and Bob be
\begin{equation}\label{eq_initial}
    \sum_{u,v=1}^d\psi_{uv}\ket{u}_{\text{S}_{\text{C}_i}}\ket{v}_{\text{S}_{\text{B}_i}},
\end{equation}
where $\ket{u}_{\text{S}_{\text{C}_i}}$ and $\ket{v}_{\text{S}_{\text{B}_i}}$ are orthonormal bases for the subsystem held by Charlie$_i$ and Bob, respectively, and $d$ is their Hilbert space dimension, which we assume to be the same without loss of generality.

After Charlie$_i$'s measurement on the $\{\ket{u}_{\text{S}_{\text{C}_i}}\}_u$ basis, the joint state of Charlie$_i$ and the $\text{S}_{\text{C}_i}\text{S}_{\text{B}_i}$ system is
\begin{equation}\label{charlie_state_meas}
\ket{\phi}_{\text{C}_i\text{S}_i}\coloneqq\sum_{u,v=1}^d\psi_{uv} \ket{u}_{\text{C}_i}\ket{u}_{\text{S}_{\text{C}_i}}\ket{v}_{\text{S}_{\text{B}_i}},
\end{equation}
where $\text{C}_i$ denotes the Hilbert space assigned to Charlie$_i$ and his lab.
To simplify the notation, we define 
\begin{equation}\psi_{c_i}\ket{c_i}_{\text{C}_i\text{S}_i}\coloneqq \sum_{v=1}^d\psi_{c_iv}\ket{c_i}_{\text{C}_i}\ket{c_i}_{\text{S}_{\text{C}_i}}\ket{v}_{\text{S}_{\text{B}_i}},
\end{equation}
so that the joint state \eqref{charlie_state_meas} can be expressed as
\begin{equation}
\ket{\phi}_{\text{C}_i\text{S}_i}=\sum_{c_i=1}^d\psi_{c_i}\ket{c_i}_{\text{C}_i\text{S}_i}.
\end{equation}
For $N$ Charlies and bipartite systems, their joint state is then
\begin{equation}\label{eq:all_charlies}
\bigotimes_{i=1}^{N}\ket{\phi}_{\text{C}_i\text{S}_i}= \!\!\!\sum_{c_1,\ldots,c_N=1}^d\left(\;\prod_{i=1}^{N}\psi_{c_i}\right)\ket{c_1,c_2,\dots,c_N}_{\rm CS}.
\end{equation}
To further simplify the notation, we denote the string representing the values of all Charlies' outcomes in each term as $\bm{c}\coloneqq c_1,c_2,\dots,c_N$. There are $d^N$ possible values of the sequence $\bm c$; let's denote them by $\bm c^{k}$, with ${k=1,\dots,d^N}$  and define ${\psi_{\bm{c}^{k}}\coloneqq\prod_{i=1}^N\psi_{c_i^k}}$. Then, \eqref{eq:all_charlies} can be expressed as
\begin{equation}
\bigotimes_{i=1}^{N}\ket{\phi}_{\text{C}_i\text{S}_i}= \sum_{k=1}^{d^N}\psi_{\bm{c}^k}\ket{\bm{c}^k}_{\rm CS}.
\label{charlies_state}
\end{equation}

Given the $N$ values of $(X,Y)$, some of the $\bm{c}^k$s will satisfy  $|f(c|xy)-f(c)|<\epsilon$, while the rest will not, where 
\begin{equation}
 |f(c|xy)-f(c)|\coloneqq\biggl|\frac{ \sum_{i=1}^N \bigl(\delta_{c_i,c}\delta_{x_i,x}\delta_{y_i,y}\bigr)}{\sum_{i=1}^N \bigl(\delta_{x_i,x}\delta_{y_i,y}\bigr)}-\frac{1}{N}\sum_{i=1}^N \delta_{c_i,c}\biggl|.
\end{equation} 
Without loss of generality, assume that the labeling of the $\bm c^k$ is such that the first $J$ terms in \eqref{charlies_state}, i.e., $\ket{\bm{c}^1}_{\rm CS}, \ket{\bm{c}^2}_{\rm CS},\ldots,\ket{\bm{c}^J}_{\rm CS}$, satisfy  $|f(c|xy)-f(c)|<\epsilon$  for all values of $c,x,y$, while the other terms do not. Then, the projector onto the eigenspace for \enquote{pass} can be written as
\begin{equation}
    \Pi_\mathrm{pass}= \sum_{k=1}^J\ketbra{\bm c^k}_{\rm CS}.
    \label{eq:projectPass}
\end{equation}

When $N\to\infty$, almost all $d^N$ strings have high Kolmogorov complexity~\cite[Sec. 2.2 and 2.4]{liIntroduction2019}, meaning that their properties are well approximated by those of a uniformly distributed random string of $d$ values. Furthermore, the statistical properties of a random subsequence of a random sequence are the same as those of the original sequence. Since the values of $(X,Y)$ in this protocol are freely chosen,\footnote{ If instead, the values of $(X,Y)$ are not freely chosen, then $J/d^N$ will not approach 1 even when ${N\to\infty}$ and thus, Veronika may not obtain the `pass' outcome and can indeed find out that $C$ and $(X,Y)$ are correlated.}  conditioning on the values of $(X,Y)$ has the effect of randomly picking a subsequence, and thus, we have $f(c|xy)\sim f(c)$. 
So, in the limit ${N\to\infty}$, we have ${J/d^N \to 1}$. 

Since the bipartite system is prepared independently from Alice's and Bob's measurement choices, the amplitudes $\{\psi_{\bm{c}^k}\}$ are by construction independent of the $N$ values of $(X,Y)$.  Together with the fact that $J/d^N\rightarrow1$,  the terms in the superposition~\eqref{charlies_state} leading to the \enquote{pass} result will dominate almost all of the amplitudes. Thus, the probability of obtaining the \enquote{pass} outcome will be arbitrarily close to 1 while the disturbance to the quantum state of all Charlies becomes arbitrarily small. Therefore, Alice and Bob can just continue the experiment after Veronika's measurement and still realize LF inequality violations as expected. 

\medskip
\textbf{A theory-independent verification protocol}

We now present our second protocol, where Veronika verifies the independence relation in a theory-independent manner, in the sense that it can be described without reference to quantum theory.

In this second protocol, Veronika talks to all the Charlies to collect their values of $C$. Then, she computes the value of $f(c|xy)-f(c)$. If Veronika obtains ${|f(c|xy)-f(c)|<\epsilon}$, she writes down her verification result as \enquote{pass}, and otherwise, she writes down her verification result as \enquote{fail}. Neither of the results she writes down contain any information about any individual values of $C$. Then, Alice proceeds with her measurement on each Charlie and his share of the bipartite system, and Bob proceeds with his measurement on each of Bob's share of the bipartite system. 

With this protocol, when $N\rightarrow\infty$, as we will show soon, it is possible to construct a realization such that it is arbitrarily likely that Veronika sees the \enquote{pass} result, and the record of this result remains publicly available. Then, assuming that Veronika is an agent capable of following the protocol, and assuming Absoluteness of Observed Events, the \enquote{pass} result gives us evidence that Eq.~\eqref{eq_CXY} holds, i.e., $P(c|xy)=P(c)$. Furthermore, it is then possible to still observe the LF inequality violations.

In order to achieve the desired violation of LF inequalities suggested by the quantum proposal, the quantum operation Alice needs to perform when this verification protocol is employed will involve an extra component compared to Alice's original measurement in the quantum proposal, so that Alice effectively undoes Veronika's interaction with the Charlies without erasing the verification record. In this quantum proposal of LF inequality violations, we also treat Veronika (denoted by V) and her verification record (denoted R) as quantum systems and assume that they are, together with Charlies, isolated from the rest of the world before Veronika sends her verification result out.

As such, we can model the interaction between Veronika and Charlies' labs (CS) as a unitary quantum process $U$ such that 
\begin{equation}
\label{eq_uni}
    U\ket{\bm{c}_k}_{\rm CS} \ket{\text{ready}}_{\rm V}=\ket{\bm{c}_k}_{\rm CS} \ket{\text{ saw \enquote{$\bm{c}_k$}}}_{\rm V},
\end{equation}
where, like in the previous protocol, $\bm c_k,k=1,\dots,d^N$ denotes the $d^N$ strings of possible values of the results of the Charlies' measurements.

Then, the state of the joint system consisting of CS, V and R after Veronika interacts with all Charlies and computes $f(c|xy)-f(c)$ can be expressed as
\begin{align}  
       &  \Biggl(\sum_{k=1}^J \psi_{\bm{c}_k}\ket{\bm{c}_k}_{\rm CS} \ket{\text{ saw \enquote{$\bm{c}_k$}}}_{\rm V}  \Biggr) \ket{\rm pass}_{\rm R} +   
              \nonumber
        \\ & \Biggl(\sum_{k=J+1}^{d^N}\!\!\!\psi_{\bm{c}_k}\ket{\bm{c}_k}_{\rm CS} \ket{\text{ saw \enquote{$\bm{c}_k$}}}_{\rm V} \Biggr) \ket{\rm fail}_{\rm R}.   
       \label{eq_hugesp}
\end{align}
Here, we again assume without loss of generality that sequences are labeled such that the first $J$ sequences of $\bm{c}_k$ yield $|f(c|xy)-f(c)|<\epsilon$ in Veronika's calculation, while for the rest $(d^N-J)$ sequences of $\bm{c}_k$, Veronika's calculation will yield $|f(c|xy)-f(c)|\geq\epsilon$. Note that here we didn't explicitly write out the systems registering the values of $X$ and $Y$ since they can be viewed as part of Veronika's memory and are thus included in V. That is, $\ket{\text{ saw \enquote{$\bm{c}_k$}}}_{\rm V}$ can be expanded into $\ket{\text{ saw \enquote{$\bm{c}_k$}}}_{{\rm V}_1}\ket{\text{ saw \enquote{$\bm{x},\bm{y}$}}}_{{\rm V}_2}$. Importantly, $\ket{\text{ saw \enquote{$\bm{x},\bm{y}$}}}_{{\rm V}_2}$ is always separable from the rest of the joint system, namely, from CS, V$_1$ and R, and thus it not necessary to explicitly account for it in this analysis.

Similarly to the first protocol, as the number of parallel runs $N$ gets larger, for combinatorial reasons the proportion of the $d^N$ sequences that present correlations with $X$ and $Y$ decreases. And since the amplitudes $\{\psi_{\bm{c}_k}\}$ are independent of the $N$ values of $(X,Y)$ by construction, the terms in the superposition where the verification record reads \enquote{pass} will dominate almost all of the amplitude.  Correspondingly, when the verification record is observed by Alice or some other agents, there will be a high probability that they will find the message to read \enquote{pass}.
When Alice reads \enquote{pass}, Alice's description of the state $CS,V$ and $R$ effectively becomes, up to normalization,
\begin{align}
\label{eq_collapse}
       \Biggl(\sum_{k=1}^J \psi_{\bm{c}_k}\ket{\bm{c}_k}_{\rm CS} \ket{\text{ saw \enquote{$\bm{c}_k$}}}_{\rm V}  \Biggr) \ket{\rm pass}_{\rm R},
\end{align}
where $J/d^N\rightarrow1$ when ${N\to\infty}$ for similar reasons given below Eq.~\eqref{eq:projectPass}.

Although the message $\ket{\rm pass}_{\rm R}$ is now separable from Veronika and the Charlies, Veronika is entangled with the Charlies. Thus, in order to continue the quantum realization of LF inequality violations, Alice, as a superobserver, needs to first apply the inverse of the unitary describing Veronika's interaction with all Charlies, i.e., $U^{\dagger}$, before she performs her quantum measurement on each of the Charlies. That is, Alice needs to have quantum control over Veronika and all Charlies. After applying $U^{\dagger}$, according to Eq.~\eqref{eq_uni}, the state of the joint system becomes  
\begin{align}
       \Biggl(\sum_{k=1,2,\ldots,J} \!\!\!\!\!\!\psi_{\bm{c}_k}\ket{\bm{c}_k}_{\rm CS}\Biggr)  \ket{\text{ready}}_{\rm V}  \ket{\rm pass}_{\rm R}.
\end{align}
Thus, the message sent out by Veronika is still publicly available, and Veronika becomes disentangled from the Charlies. When $N\rightarrow\infty$, the state of all Charlies will be arbitrarily close to the one before Veronika did her verification. As such, Alice and Bob can now proceed with their measurements for realizing LF inequality violations as in the original quantum proposal.

Note that although in the above quantum analysis, the validity of quantum theory is assumed and the verification protocol is paired with an extra quantum operation performed by Alice, i.e, the inversion of Veronika's interaction with Charlies, the description of the protocol itself is theory-independent and only relies on Veronika being a trustworthy agent who can communicate with the Charlies and compute relative frequencies. The quantum analysis and the need for Alice to perform $U^{\dagger}$ is only to show that such a verification protocol will not prevent the quantum realization of LF inequality violations.

\section{Conclusions}
\label{sec_conc}

In this work, we introduced two new perspectives to the LF no-go theorem~\cite{bongStrong2020}: the first recasts LF inequalities as special cases of monogamy relations, the second recasts LF inequalities as causal compatibility inequalities associated with a \emph{nonclassical} causal marginal problem. Both perspectives bring new tools and insights to further elucidate the underlying landmark result, and illuminate new potential avenues for future research.

By casting the LF inequalities as monogamy relations, we see that these inequalities arise precisely because of the correlation that Alice must occasionally share with Charlie,  provided that Eqs.~\eqref{eq_LA}---a statistical constraint analogous to the no-signaling condition in a tripartite Bell scenario---is satisfied.
This point of view links the study of extended Wigner's Friend scenarios with ideas previously leveraged in quantum information technologies such as quantum key distribution and randomness amplification. This connection now enables researchers to share technical toolboxes originally developed for distinct motivations. More speculatively, this connection may also give hints on new no-go theorems in extended Wigner's friend scenarios, for example, by looking at monogamy relations derived from constraints other than no-signaling.

Note that the LF monogamy relations are derived in a manner agnostic to whatever particular physical theory may be governing the systems shared by the agents. The theory-independent nature of their derivation means there is no escaping the LF monogamy relations by modifying the causal explanation to employ quantum, GPT, or other types of nonclassical systems as common causes. This is formalized by our results from the causal compatibility perspective, using the framework of $d$-sep causal modeling we defined in Sec.~\ref{sec_preliminaries}.

Indeed, while one \emph{can} explain Bell inequality violations by invoking a \emph{quantum} common cause in the Bell DAG, we showed by contrast that associating GPT (or even more exotic) systems with the latent node of the LF DAG is \emph{useless} for explaining LF inequality violations.
This follows from the fact that the LF inequalities are causal compatibility inequalities stemming from a \textit{nonclassical} causal marginal problem; see Theorem~\ref{thm_LF_DAG}. This is 
related to the fact that the LF no-go theorem can be seen as stronger than Bell's theorem
~\cite{bongStrong2020,cavalcantiImplications2021}.  We refer the reader to Appendix~\ref{sec_comparison} for a comparison of the probability distributions compatible with the Bell DAG versus the LF DAG under different theory prescriptions and constraints.

Besides showing that the LF inequalities are nonclassical causal compatibility inequalities of the LF DAG---i.e., that the LF inequalities follow from the LF DAG with any theory prescription under $d$-sep causal modeling---we also showed (Theorem~\ref{thm_only})  that the LF DAG, under GPT causal modeling, does not impose any extra constraints on compatible probability distributions beyond those imposed by the metaphysical assumptions in previous formulations of LF no-go theorems. Since the LF DAG with GPT causal modeling constitutes the \emph{most powerful} causal explanation satisfying the fundamental causal principles associated with the LF causal-metaphysical assumptions (see Theorem~\ref{thm_RCA} and Corollary~\ref{corollary_relativistic}), one may interpret the LF DAG as a visual and memorable representation of the LF theorem. Mnemonically, the LF DAG with GPT or $d$-sep causal modeling stands in place of the LF causal-metaphysical assumptions.

\medskip

Our results also underscore how Extended Wigner's Friend scenarios pose formidable challenges for the field of nonclassical causal inference. First, we strengthened the challenge previously presented in Ref.~\cite{cavalcantiImplications2021} (Theorem~\ref{thm_RCA}); that is, we showed that not \emph{only} quantum causal models, but rather \emph{all} $d$-sep causal models are in tension with \RCA\ and \IS. Second, our Theorem~\ref{thm_nft} presents a new conundrum for the field: \textit{no} causal structure under \textit{non}classical causal modeling can explain LF inequality violations without violating the principle of No Fine-Tuning. Notably, Theorem~\ref{thm_nftcomp} further extends Theorem~\ref{thm_nft} beyond \emph{acyclic} causal structures to most (and potentially all) \textit{cyclic} ones. To show this, we developed a generalization of $d$-sep causal modeling---\emph{compositional causal modeling}---adapting the No Fine-Tuning principle to cyclic causal structures, as explained in Sec.~\ref{sec_cyc}.
 
Our results mean that the most general available framework for causal reasoning that assumes \AOE\ cannot account for the violations of the LF inequalities without simultaneously compromising 1) either \RCA\ or \IS,  and also 2)  
the No Fine-Tuning principle of causal discovery. This is a radical finding, as this framework is already extremely permissive: it does not constrain the specific theory that can be used in the causal explanation, and can thus accommodate even certain exotic theories beyond GPTs; it also allows cyclic causal models, as long as the separation rule is compositional, such as the $\sigma$-separation rule~\cite{forre2017markov} or the $p$-separation rule~\cite{Carla2023}. 

Speculatively, how might the existing causal modeling framework be generalized even further, so as to be able to sensibly account for LF inequality violations? We see two divergent possibilities: one is to develop a framework for \emph{non}compositional causal modeling, the other is to reject \AOE\ by, e.g.,  generalizing the definition of a causal structure and how its nodes are related to observed events.

We are skeptical that the first option could be fruitful. Firstly, while noncompositional graphical-separation rules for cyclic causal models might salvage the No Fine-Tuning principle, it cannot help with the fact that  \RCA\ and \IS\ would imply \emph{acyclic} causal structure and thus render cyclic causal models irrelevant. Additionally, noncompositionality would imply that causal dependence relations of a set of variables on another set do not reduce to the individual causal dependence of each element of the former on its causes. These \emph{holistic} properties would nevertheless be causally related to other variables but presumably not themselves appear as nodes in the graphical model. Thus, one can ask in what sense the model provides a causal explanation for these holistic properties at all.
Finally, note that a noncompositional separation rule that implies conditional independence relations \emph{beyond} those implied by $d$-separation could be problematic, since it would not reduce to the $d$-separation rule for DAGs. After all, even classical causal models can violate every conditional independence relation \emph{not} implied by a $d$-separation relation in a DAG. 

Perhaps a more fruitful way forward is the second possibility: to generalize the definition of causal structures and how they are associated with observed events. So far, the causal structure used in the framework is single and static, and all observed events are associated with a single node that carries a classical random variable; \AOE, in other words, is embedded in the compositional causal modeling frameworks. If one denies the absolute nature of Charlie's outcome, the premise of the marginal problem would no longer hold: there will no longer be a well-defined joint probability distribution $P(abc|xy)$ simultaneously having $\{P(ab|xy), P(ac|x{=}1)\}$ as its marginals. 
Some interpretations of quantum theory deny \AOE. For example, in Relational Quantum Mechanics~\cite{rovelli2021relational,dibiagioStable2021}, QBism~\cite{fuchs2016qbism}, and Everettian quantum mechanics~\cite{wallace2014emergent}, observed events may be relative to specific systems, agents, and branches, respectively.\footnote{ These interpretations have also been argued to be able to resolve other Wigner's-friend-type arguments such as the Frauchiger-Renner argument~\cite{Frauchiger2018}, which does not explicitly assume \AOE. However, the assumptions in the Frauchiger-Renner argument~\cite{Frauchiger2018} have not been formulated in a causal manner, and so we have not analyzed it here.}  However, it is unclear how to construct causal models in such interpretations. Recently, proposals for developing causal modeling in the absence of \AOE\ have been discussed in, e.g., ~\cite{ormrod2024quantuminfluenceseventrelativity,vilasini2022generalframeworkconsistentlogical}.

\medskip

The realization that Bell inequalities are classical causal compatibility inequalities has inspired the exploration of novel scenarios exhibiting quantum advantages and the construction of frameworks for modeling nonclassical causality. Analogously, we hope that our work will foster collaborations between two previously disconnected communities, namely, the community studying extended Wigner's friend theorems and the one focusing on nonclassical causal inference. For example, just as causal compatibility \emph{inequalities} (instead of equalities) had been much overlooked before the work of Ref.~\cite{woodLesson2015}, so too \emph{causal marginal problems} have thus far received relatively sparse consideration by the classical causal inference community, let alone in nonclassical contexts. We anticipate that the techniques and findings in \emph{nonclassical causal inference for marginal problems} will ultimately expose novel insight-rich scenarios related to (or generalizations of) the extended Wigner's friend scenario. Finally, as mentioned, we hope that our work will inspire both communities to find a meaningful further generalization of the compositional causal modeling framework.

\begin{acknowledgements}
This project was originated in the discussions YY, MMA and ADB had at the 2022 Kefalonia Foundations workshop. Many thanks to V. Vilasini for sharing with us many valuable insights on cyclic causal models. We further thank Robert W. Spekkens, Howard M. Wiseman, John Selby, Veronika Baumann, Marcin Paw\l{}owski, Marc-Olivier Renou, Victor Gitton, Ilya Shpitser, Isaac Smith, Markus M{\"u}ller, Caroline L. Jones, Eleftherios Tselentis and Roberto D. Baldijao for insightful discussions. 

YY, MMA and EW were supported by Perimeter Institute for Theoretical Physics. Research at Perimeter Institute is supported in part by the Government of Canada through the Department of Innovation, Science and Economic Development and by the Province of Ontario through the Ministry of Colleges and Universities. YY and MMA were also supported by the Natural Sciences and Engineering Research Council of Canada (Grant No. RGPIN-2017-04383). ADB was supported by the John Templeton Foundation (Grant ID\#61466) as part of the ``Quantum Information Structure of Spacetime (QISS)'' project (\hyperlink{http://www.qiss.fr}{qiss.fr}). DS was supported by the National Science Centre, Poland (Opus project, Categorical Foundations of the Non-Classicality of Nature, project no. 2021/41/B/ST2/03149).  EGC and YY were supported by grant number FQXi-RFP-CPW-2019 from the Foundational Questions Institute and Fetzer Franklin Fund, a donor-advised fund of Silicon Valley Community Foundation, and by the Australian Research Council (ARC) Future Fellowship FT180100317. 
\end{acknowledgements}

\appendix

\section{Additional causal inference background}

\subsection{GPT-compatibility}
\label{app_GPTcom}

The following is adapted from \cite[Section 3]{hensonTheoryindependent2014}, which uses the framework of OPTs~\cite{ChiribellaProbabilistic2010}. We assume that the reader has basic familiarity with the GPT framework~\cite{Muller_2021,Plavala_2023}, which we use instead. For our purpose, the difference between GPTs and OPTs is not important and all of our results apply to both.

A probability distribution over observed nodes, denoted as $P(\obs(\mathcal{G}))$, is GPT-compatible\footnote{Sometimes it is also said to satisfy the \enquote{generalized Markov condition}.~\cite{hensonTheoryindependent2014}} with a given causal structure $\mathcal G$ if and only if it can be generated by a GPT in the way prescribed by $\mathcal G$, namely, if and only if there exists a GPT $\mathsf T$ such that:
\begin{itemize}[-]
    \item a system in $\mathsf T$ is associated to each edge that starts from a latent node in $\mathcal G$,
     \item for each latent node $Y$ and for any value of its observed parents, denoted $\Opa(Y)$, there is a channel  $\mathcal C_Y(\opa(Y))$ in $\mathsf T$ from the systems associated with latent-originating edges incoming to $Y$ to the composite system associated with all edges outgoing from $Y$,
    \item for every value $x$ of an observed node $X$, and for any value $\opa(X)$ of its observed parents $\Opa(X)$, there is an effect $\mathcal E_X(x|\opa(X))$ in $\mathsf T$ on the system composed of all systems associated with edges incoming in $X$, such that $\sum_x \mathcal E_X(x|\opa(X))$ is the unique deterministic effect on the systems associated to edges coming from latent nodes to $X$,
    \item $P(\obs(\mathcal{G}))$ is the probability obtained by wiring the various tests $\mathcal E_X(x|\opa(X))$ and channels $\mathcal C_Y(\opa(Y))$.
\end{itemize}
Note that the theory $\mathsf T$ may contain several system types, e.g., including also classical ones.
According to these definitions above, an observed node with no latent parents is simply associated with a (conditional) probability distribution (a set of effects on the trivial system that sum to the unique deterministic effect on the trivial system), while a latent node with no latent parents is associated to a classically-controlled state over the relevant systems (i.e., a controlled deterministic channel from the trivial system).

GPT causal modeling reduces to classical causal modeling when all the nodes in the causal structure are observed nodes. In that case, the probability distribution over all the variables of $\cal G$ must be in the form of
\begin{equation}
\label{eq:Markov1}
P\left(\obs(\mathcal{G})\right)=\prod_{N} P\left(n|\pa(N)\right).
\end{equation}
Another common special case of GPT causal modeling is quantum causal modeling, where systems associated to edges outgoing from latent nodes may be either classical or quantum.  For example, when the systems  outgoing from the latent common cause of the Bell DAG are all quantum, the probability distribution over all observed variables of $\cal G$ must be in the form
\begin{equation}
    P(abxy)=\Tr[\rho_{AB}(E_{a|x}\otimes F_{b|y})]P(x)P(y),
\end{equation}
where $\rho_{AB}$ is the state of some quantum state associated with the latent node, and $E_{a|x}$ and $F_{b|y}$ are POVMs associated with nodes $A$ and $B$, respectively.

\subsection{The LF DAG as a GPT diagram}

Let us apply the definition of GPT-compatibility to the LF DAG.

First, we note that there is only one latent node $\mathcal L$, with three outgoing edges, and no parents. Therefore, $\mathcal L$ will be associated with a tripartite state preparation $\rho$. $\mathcal L$ has three observed children, $A,B,$ and $C$, and each of them has $\mathcal L$ as their only latent parent. This means that each of them will be associated with a family of effects on one of the three systems coming out of $\rho$. $C$ has no other parents, so it is associated with a family $\{\mathcal E_C(c)\}$, $B$ has only one other parent, so it is associated with $\{\mathcal E_B(b|y)\}$, and $A$ has two observed parents, so it corresponds to $\{\mathcal E_A(a|cx)\}$. Finally, the two observed nodes $X$ and $Y$ are associated with probability distributions $P(x)$ and $P(y)$, as they have no parents. Note that the three systems coming out of $\rho_\mathcal{L}$ do not have to be of the same kind. 

Thus, a probability distribution $P(abcxy)$ is GPT-compatible with the LF DAG if it corresponds to a diagram with the following shape
\begin{equation}
   \begin{tikzpicture}
	\begin{pgfonlayer}{nodelayer}
		\node [style=state] (0) at (0, -1.75) {$\rho_\mathcal{L}$};
		\node [style=effect] (3) at (-3.5, 2.25) {\raisebox{1ex}{$\mathcal E_A${\small$(a|xc)$}}};
		\node [style=diamond] (4) at (4.5, 0) {$P(y)$};
		\node [style=diamond] (5) at (-4.5, 0) {$P(x)$};
		\node [style=effect] (6) at (0, 1.25) {\raisebox{1ex}{$\mathcal E_C${\small$(c)$}}};
		\node [style=effect] (7) at (3.5, 2.25) {\raisebox{1ex}{$\mathcal E_B${\small$(b|y)$}}};
		\node [style=none] (8) at (-1, -1.75) {};
		\node [style=none] (9) at (1, -1.75) {};
	\end{pgfonlayer}
	\begin{pgfonlayer}{edgelayer}
		\draw [style=black, in=-90, out=90] (8.center) to (3);
		\draw [style=black] (0) to (6);
		\draw [style=black, in=-90, out=90] (9.center) to (7);
	\end{pgfonlayer}
\end{tikzpicture}
.
\end{equation}
One can write this equivalently as the formula
\begin{equation}
    \!\!P(abcxy)\!=\!P\big[\mathcal E_A(a|cx)\mathcal E_C(c)\mathcal E_B(b|y)\rho_\mathcal{L}\big]P(x)P(y),
\end{equation}
where the values of $P\big[\mathcal E_A(a|cx)\mathcal E_C(c)\mathcal E_B(b|y)\rho_\mathcal{L}\big]$ will be specified by the GPT. To derive the LF inequalities, there is an extra condition on $\mathcal E_A(a|cx)$, namely that $\mathcal E_A(a|c,x{=}1)$ is $\delta_{a,c}$ times the unique deterministic effect.

\subsection{$d$-separation and its compositionality}
\label{appendix_dsep}

In this appendix, we present the $d$-separation rule developed in Ref.~\cite{Geiger1988,Geiger1990} and the compositionality of $d$-separation relations.

In what follows, a \emph{path} of nodes in a DAG is a sequence of nodes connected by arrows, independently of the way these arrows are oriented.

\begin{definition}[Blocked path]
    Let $\cal G$ be a DAG, $p$ be a path of nodes in $\mathcal{G}$ and $\bm{Z}$ be a set of nodes of $\mathcal{G}$. We say that the path $p$ is \emph{blocked by the set} $\bm{Z}$ if at least one of the following hold:
    \begin{enumerate}
        \item  $p$ includes a sequence $A\rightarrow C \rightarrow B$, called a \emph{chain}, where $C\in \bm{Z}$. 
        \item $p$ includes a sequence $A\leftarrow C \rightarrow B$, called a \emph{fork}, where $C\in \bm{Z}$. 
        \item $p$ includes a sequence $A\rightarrow C \leftarrow B$, called a \emph{collider}, where $C\not\in \bm{Z}$ and $D\not\in \bm{Z}$ for all the descendants $D$ of the node $C$. 
    \end{enumerate}
\end{definition}
When a path $p$ has a collider $A\rightarrow C \leftarrow B$, we will say that $p$ includes a \emph{collider on $C$}.

\begin{definition}[$d$-separation relation]
    \label{def_dsep}
    \hspace{1em}\\\noindent{\normalfont\textbf{[Singleton-pair separation criterion]}} Let $\cal G$ be a directed graph, where $X$ and $Y$ are single nodes in $\cal G$, and $\boldsymbol{Z}$ indicates a \emph{set} of nodes of $\cal G$. We say that  $X$ and $Y$ are \emph{$d$-separated by the set} $\boldsymbol{Z}$, denoted ${X\perp Y|\boldsymbol{Z}}$, if all of the paths connecting $X$ to $Y$ are blocked by $\bm{Z}$. \par
    
    \noindent{\normalfont\textbf{[Setwise separation criterion]}} Two \emph{sets} of nodes $\boldsymbol{X}$ and $\boldsymbol{Y}$ are said to be $d$-separated by $\boldsymbol{Z}$, denoted ${\boldsymbol{X}\perp \boldsymbol{Y}|\boldsymbol{Z}}$, if and only if every node in $\boldsymbol{X}$ is $d$-separated from every node in $\boldsymbol{Y}$ given $\boldsymbol{Z}$.
\end{definition}
Note that the relationship between setwise $d$-separation and singleton-pair $d$-separation implies that $d$-separation is compositional in the sense of Definition~\ref{def_comp}. The compositionality of $d$-separation was previously pointed out in Ref.~\cite{Koster99} and is also emphasized in Ref.~\cite{sadeghi2017faithfulness}.

As we saw in Section~\ref{sec_preliminaries}, the d-separation \textit{rule} states that, whenever a DAG $\mathcal{G}$ satisfies a $d$-separation relation ${\boldsymbol{X}\perp \boldsymbol{Y}|\boldsymbol{Z}}$ over observed nodes $\boldsymbol{X}, \boldsymbol{Y},\boldsymbol{Z}$, a probability distribution $P$ needs to satisfy the conditional independence ${P(\boldsymbol{X}\boldsymbol{Y}|\boldsymbol{Z})=P(\boldsymbol{X}|\boldsymbol{Z})P(\boldsymbol{Y}|\boldsymbol{Z})}$ in order to be compatible with $\mathcal{G}$, under all nonclassical theory prescriptions. 

By way of contrast, note that conditional independence relations are \emph{not} compositional. For instance, suppose that in a certain probability distribution $P$, the variable $U$ is independent of the variable $W$ and the variable $V$ is also independent of the variable $W$. This \emph{does not imply} that the joint variable $UV$ is independent of $W$: for example, it is possible that $W$ determines the parity of $U$ and $V$, even if it is independent of each one individually. 

\subsection{Cyclic models violating the $d$-separation rule}
\label{appendix_cyclic}

In this appendix, we will use an example taken from Ref.~\cite{forre2017markov} to explain why the $d$-separation rule is not always valid for cyclic causal structures. In this example, there are no latent nodes; since observed nodes are always classical, the theory prescription does not make a difference here. The example violates both the $d$-separation rule and the $\sigma$-separation rule, but obeys the $p$-separation rule. A counterexample to the $p$-separation rule is not known yet, but it \emph{is} known that no such counterexample can be found within quantum causal models.

\begin{figure}[h]
    \centering
	\includegraphics[width=0.12\textwidth]{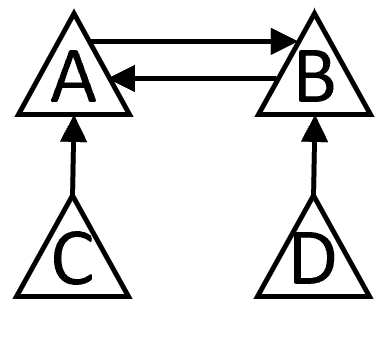}
	\caption{Cyclic directed graph that presents the $d$-separation relation $C\perp D$ but is classical-compatible with distributions where $C$ and $D$ are correlated.} 
	\label{fig_cyclic_example2}
\end{figure}

The example concerns the causal structure of Fig.~\ref{fig_cyclic_example2}. In it, there is no open path between $C$ and $D$, so this graph satisfies the $d$-separation relation $C\perp D$. In fact, it also satisfies the $\sigma$-separation relation $C\oldperp_{\sigma} D$, as shown in Ref.~\cite{forre2017markov}. However, it is possible to construct a functional model on this graph where $C$ and $D$ are (perfectly) correlated. 

Such a model is obtained from the functions ${a=c\oplus b}$  and ${b=a\oplus d}$, where all the variables are binary and $\oplus$ indicates summation modulus two. Let us consider all the possible cases:
\begin{itemize}
    \item When $c=d=0$, we have $a=b$;
    \item When $c=d=1$, we have $a=b\oplus 1$;
    \item When $c\neq d$, there is no solution for $A$ and $B$.
\end{itemize}
That is, the functional model only has solutions when ${c=d}$, so $c$ and $d$ are (perfectly) correlated.

Thus, this is an example where both the $d$-separation rule and the $\sigma$-separation rule are not valid.

\section{Comparing the LF DAG and the Bell DAG}
\label{sec_comparison}

In this section, we use causal modeling reasoning to derive and provide intuitions for various relationships between different sets of probability distributions related to Bell's theorem and the LF theorem.

\begin{figure}[h!]
    \centering
    \includegraphics[width=0.47\textwidth]{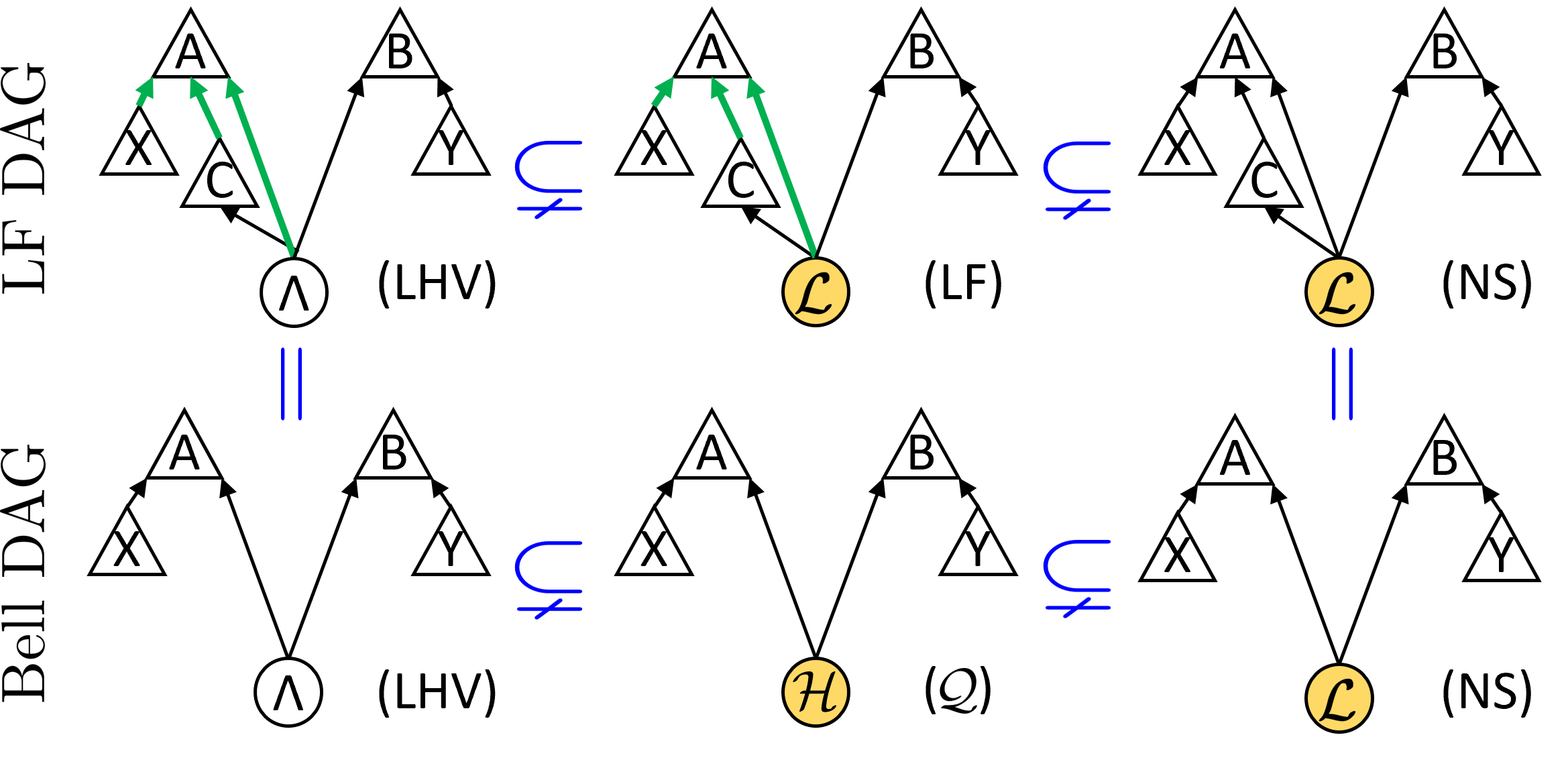}
    \caption{The set symbols indicate the relations between the sets of correlations $P(ab|xy)$ that are also marginals of compatible $P(abc|xy)$, which must satisfy $P(a|c,x{=}1)=\delta_{a,c}$ when the green arrows are present. The names of these sets are in parenthesis and correspond to the respective names of the sets of correlations depicted in Fig.~\ref{fig_polytope}. A latent node labeled by $\Lambda$ can only be associated with a classical random variable, while $\cal L$ may be associated with any GPT systems and $\cal H$ (stands for Hilbert spaces) is associated with any quantum systems. }
    \label{fig_compare}
\end{figure}

\begin{figure}[h!]
    \centering
    \includegraphics[width=0.3\textwidth]{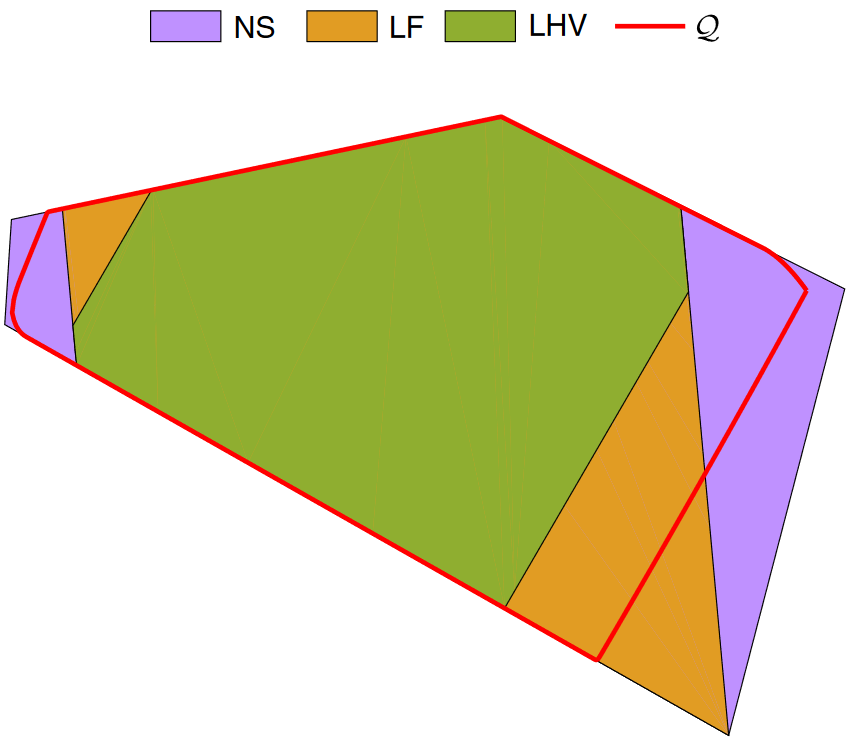}
    \caption{A two-dimensional slice of the Local Hidden Variable (LHV) polytope, the LF polytope, the No-Signaling (NS) polytope, and the boundary of quantum ($\cal Q$) correlations, for a particular LF scenario. See Ref.~\cite{bongStrong2020} (from which the figure is reproduced) for further details. }
    \label{fig_polytope}
\end{figure}

Fig.~\ref{fig_compare} shows the relationship between the LF DAG and the Bell DAG. Although we have been talking about the LF \emph{monogamy relations} throughout most of this text, this figure refers to the special case of the LF \emph{inequalities}; that is, whenever a DAG has green arrows in Fig.~\ref{fig_compare}, we assume ${P(a|c,x{=}1)=\delta_{a,c}}$. The inclusion signs between the DAGs indicate the relationships between the sets of distributions $P(ab|xy)$ that can be explained by them. Fig.~\ref{fig_compare} also labels each of these sets with the name of the corresponding sets of correlations depicted in Fig.~\ref{fig_polytope}.

\begin{table*}[!ht]
\footnotesize
    {\centering
    \begin{tabular}{l | p{4.7cm} p{.3cm} |p{5.7cm} p{0.8cm}}
    \hline
    &Bell inequalities & &LF inequalities &\\ \hline
     DAG &     $$\includegraphics[width=0.1\textwidth]{Bell.png}$$&
    & $$\includegraphics[width=0.1\textwidth]{LF.png}$$&\\ \hline
        \multirow{3}{2cm}{Constraints implied by the DAG under $d$-sep causal modeling}& $P(ab|xy)     =\sum_{\lambda}P(\lambda|xy)P(ab|xy\lambda)$ &   
        & $P(ab|xy)
        =\sum_{c}P(c|xy)P(ab|cxy)$ & 
        \\ \cline{2-5}
        & $\Lambda \perp XY$ \newline $\Rightarrow P(\lambda|xy)=P(\lambda)$ &
        & $C\perp XY$ \newline $\Rightarrow P(c|xy)=P(c)$ &
        \\  \cline{2-5}
        & $AX\perp BY|\lambda$ \newline $\Rightarrow P(ab|xy\lambda)=P(a|x\lambda)P(b|y\lambda)$ & 
        & $AC\perp Y|X\quad BC\perp X|Y$   \newline $\Rightarrow P(ac|xy)=P(ac|x), P(bc|xy)=P(bc|y)$ 
        \\
        \hline
        Additional constraint& None & &${P(ac|x{=}1)=\delta_{a,c}P(c|x{=}1)}$  
        \\ \hline
    \end{tabular}
    \caption{\textbf{Comparison of the derivation of Bell inequalities using the Bell DAG and the derivation of LF inequalities using the LF DAG.} }
    \par
    \label{tab:compare}}
\end{table*}

Let us explain each set relation in Fig.~\ref{fig_compare}.

\emph{The left equal sign:} For the LF DAG with a classical latent node, since we are only interested in $P(ab|xy)$, we can merge $C$ and the latent node $\Lambda$ into a new latent node $\Lambda'=\{\Lambda,C\}$. The constraint that $P(a|c,x{=}1)=\delta_{a,c}$ is translated into a condition on the possible dependence of $A$ on $\Lambda'$: $A$ has to be a copy of a part of $\Lambda'$ (namely, $C$) when $x=1$. However, this condition \emph{does not} impose any restriction on the dependence of $A$ on $\Lambda'$: being a copy of an arbitrary part of a latent node is equivalent to having an arbitrary dependence on the latent node as a whole. Therefore, if treated as a classical causal structure, the LF DAG is compatible with the same sets of $P(ab|xy)$ as the classical Bell DAG, even when demanding ${P(a|c,x{=}1)=\delta_{a,c}}$.

\emph{The right equal sign:} Since we are only concerned with $P(ab|xy)$ and we do not demand $P(a|c,x{=}1)=\delta_{a,c}$ for the upper DAG (there is no green arrow), the node $C$ can simply be incorporated into the GPT latent node $\mathcal{L}$. 

\emph{The top left strict inclusion sign:} The constraint $P(a|c,x{=}1)=\delta_{a,c}$ only applies to the case when $x=1$. Thus, for example, where $x$ can take at least three different values, namely, $x=1$, $2$, or $3$, the distributions $P(ab|x=2,y)$ and $P(ab|x=3,y)$ compatible with the upper middle diagram can violate the CHSH inequalities when the common cause is quantum. On the other hand, the distributions compatible with the upper left diagram satisfy Bell inequalities, as made explicit by the equal sign between the leftmost DAGs.

\emph{The top right strict inclusion sign:} From the right equal sign, we know that any $P(ab|xy)$ satisfying no-signaling is compatible with the LF DAG when $\mathcal{L}$ is described by an arbitrary GPT. Thus, there exists $P(ab|xy)$ compatible with the LF DAG violating LF inequalities. The problem is that none of them can do so while satisfying the constraint $P(a|c,x{=}1)=\delta_{a,c}$. 

The relations in the second row are well-known~\cite{Brunner2014}. It shows how quantum theory is more powerful than classical theory but is not the most powerful GPT in explaining Bell inequality violations. 

The fact that the set corresponding to the top-middle diagram is neither a subset nor a superset of the set for the bottom-middle diagram reflects the fact, proven in Ref.~\cite{bongStrong2020} and displayed in Fig.~\ref{fig_polytope}, that the quantum ($\cal Q$)  set of correlations neither contains nor is contained in the LF polytope. 

One important thing to note is that the quantum set of empirical correlations ($\cal Q$) in an LF scenario is exactly the set of correlations $P(ab|xy)$ compatible with the Bell DAG under the quantum theory prescription, i.e., it is the set represented by the bottom-middle diagram in Fig.~\ref{fig_compare}. This is because, in the quantum protocol, we assume that Alice can rewind Charlie's measurement so that when $x\neq 1$, Alice ends up measuring Charlie's share of the entangled system directly. That is, regardless of $x=1$ or $x\neq 1$, the outcome obtained by Alice can be viewed as an outcome of a measurement on Charlie's share of the entangled system. However, the fact that the quantum set of correlations ($\cal Q$) is exactly the set of correlations $P(ab|xy)$ compatible with the Bell DAG under quantum theory prescription does \emph{not} mean that the bottom-middle diagram represents a good causal explanation for the minimal LF scenario. This is so because it would not explain Charlie's experience as an observer.

The comparison between the derivation of Bell inequalities and LF inequalities using their respective DAGs is shown in Table~\ref{tab:compare}.

\section{An example of fine-tuned models: the cryptographic protocol}
\label{appendix_cypher}

A causal model for a cryptography protocol will be fine-tuned. Nevertheless, this fine-tuning is not a problem, since it can be further explained by how the cryptographer carefully designs the encryption algorithm to wash out correlations.

\begin{figure}[htbp]
    \centering
    \includegraphics[width=0.11\textwidth]{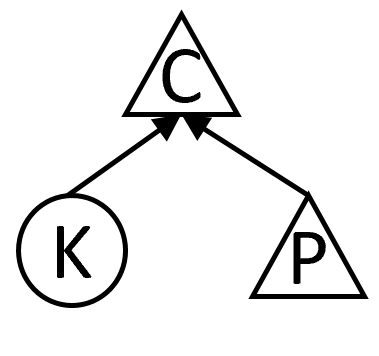}
    \caption{\textbf{The DAG representing the causal structure for the cryptographic protocol.} $P$ represents the plaintext, $K$ represents the key and $C$ represents the cyphertext.}
    \label{fig_cypher}
\end{figure}

To understand this, consider a cryptography protocol where $P$ represents the plaintext (the message to be sent), $K$ represents the key and $C$ represents the cyphertext (the encoded message). The cyphertext is constructed from the plaintext and the key, so the causal structure of this process is the one given in Fig.~\ref{fig_cypher}.

Assume that an agent has access to the plaintext and the cyphertext, but not the key. Thus $K$ is a latent node in Fig.~\ref{fig_cypher}, while $C$ and $P$ are observed nodes.

The encryption protocol is designed to make the cyphertext independent of the plaintext. Take the example where $C$, $P$ and $K$ are all binary variables, and $C$ responds to its parents as $c=p \bigoplus k$, where $\bigoplus$ indicates sum modulus two. If $K$ is uniformly distributed between the values $0$ and $1$, then $C$ will be uniformly distributed as well, independently of the value of $P$. This shows that $P$ is statistically independent of $C$, even though there is a causal influence from $P$ to $C$. Therefore this causal model, which is clearly the correct model for this phenomenon, is fine-tuned.

Note, however, that the causal structure of the cryptographic protocol can \emph{in principle} be determined from observations and interventions, if we also know the value of the latent variable (in this case, the key). That is, if we have access to the value of the key, by varying the plaintext we can attest that the plaintext indeed has causal influence on the cyphertext. This is in contrast with the case of the fine-tuned causal models for the Bell and LF scenarios, where (assuming it is impossible to signal faster than light) it is \emph{in principle impossible} to 
carry out arbitrary interventions (such as observations) on the latent nodes (i.e., the hidden variables). Consequently in these cases, it is impossible to single out one causal structure in the same way that one can do for the cryptographic example.

\section{Discussions of the verification protocol in Sec.~\ref{sec_vero}}
\label{app_nft}

Here, we list a few possible concerns that one may have regarding the verification protocols presented in Sec.~\ref{sec_vero}, and provide our analysis. Note that as mentioned in the main text, the validity of our No Fine-Tuning theorems (Theorem~\ref{thm_nft} and \ref{thm_nftcomp}) does not rely on the verification protocols.

\medskip

\textbf{Why not use the following alternative verification protocol?}

Consider the following protocol. The verifier randomly halts the experiment after Alice selects her measurement choice but before she executes the measurement. Subsequently, the data of $C$, $X$, and $Y$ in these runs can be employed to examine the interdependence among these variables.  This is similar to the cross-validation approach often used in classical causal inference.

We favor the Veronika protocols detailed in Sec.~\ref{sec_vero}. This preference arises from the fact that in the protocol in the previous paragraph, all values of $C$, $X$, and $Y$ used for verification are no longer causally linked to Alice's measurement. Moreover, in those runs, it is impossible to continue the quantum experiment because Alice cannot reverse Charlie's measurement when $C$ is known to a classical verifier. Essentially, this protocol cannot establish independence relations in runs involving Alice's reversal of Charlie's measurement. In other words, this protocol hinges on the runs terminated by the verifier for verification as being a  representative sample of all runs of the experiment. In contrast, the Veronika protocols gather data for establishing independence from \emph{all} experiment runs, even when Alice reverses Charlie's measurement subsequently.

\medskip

\textbf{The Veronika protocols cannot be done in the spacelike-separated LF scenario.}

Since Veronika needs to collect data for both $X$ and $Y$ in conjunction with $C$, and given that Alice must reverse the potential entanglement between Veronika and $C$ before proceeding with her measurement, $A$ can no longer be spacelike-separated from $Y$.

However, this is of no concern for the No Fine-Tuning no-go theorems (Theorem~\ref{thm_nft} and~\ref{thm_nftcomp}) since the spacetime constraint is irrelevant there. The No Fine-Tuning no-go theorems do not assume \RCA.

Besides, it is possible to modify the Veronika protocols even in case of spacelike separation, but with an extra assumption motivated by the fact that Charlie informs Alice of his measurement outcome when $x=1$. This assumption is that the causal structure must have an arrow from $C$ to $A$. Under this assumption, No Fine-Tuning with Eq.~\eqref{eq_signaling} and ${P(c|x)=P(c)}$ implies that $C$ and $Y$ must be $d$-separated. This is because, if they were not, then the arrow from $C$ to $A$ would imply $A\not\oldperp_\text{d} Y|X$ (when there is a path from $C$ to $Y$ unblocked by $X$) and/or $C\not\oldperp_\text{d} X$ (when there is a path from $C$ to $Y$ blocked by $X$). However, No Fine-Tuning with Eq.~\eqref{eq_signaling} and ${P(c|x)=P(c)}$ demands that $A\oldperp_\text{d} Y|X$ and  $C\oldperp_\text{d} X$. Therefore, $C$ and $Y$ must be $d$-separated. As such, Veronika only needs to verify the independence of $C$ from $X$ instead of also from $Y$; in this case, Alice can still make her measurement in a spacelike-separated manner from Bob's measurement.

\medskip

\textbf{The Veronika protocols may be susceptible to the ``simultaneous memory loophole''.}

Readers who are familiar with loophole-free Bell tests may worry that performing the runs of an LF experiment in parallel may lead to the so-called ``memory loophole'' and in particular the ``simultaneous memory loophole''~\cite{Barrett_2002}:
carrying runs in parallel may increase the likelihood of correlations between runs and hence lead to a higher chance of violating the inequalities.

This worry is essentially about the assumption that all the data in the parallel runs are independent and identically distributed. Such an assumption is also needed when the runs of the experiments are carried out sequentially instead of in parallel (and hence the \enquote{memory loophole} still exists in sequential experiments). It has been argued in the context of Bell experiments that it is possible to close the simultaneous loophole by careful experimental designs and sophisticated statistical analysis~\cite{Barrett_2002,jpa512003bib22,Bierhorst_2015}.
This suggests that there may also be a way to close the simultaneous memory loophole in the Veronika protocols.

However, our case is not completely analogous, as there is an additional potential source of correlations arising from Veronika's interactions with all the Charlies, which may introduce correlations between the runs. We leave it to future research to understand what kind of additional simultaneous memory loophole it will bring and how to resolve it.

\bibliographystyle{quantum}
\bibliography{reference.bib}

\begin{thebibliography}{10}

\bibitem{glymourMarkov2006}
Clark Glymour.
\newblock ``{Markov Properties and Quantum Experiments}''.
\newblock In William Demopoulos and Itamar Pitowsky, editors, Physical
  {{Theory}} and Its {{Interpretation}}: {{Essays}} in {{Honor}} of {{Jeffrey
  Bub}}.
\newblock \href{https://dx.doi.org/10.1007/1-4020-4876-9\_5}{Pages 117--126}.
\newblock The {{Western Ontario Series}} in {{Philosophy}} of {{Science}}.
  {Springer Netherlands}~(2006).

\bibitem{woodLesson2015}
Christopher~J Wood and Robert~W Spekkens.
\newblock ``{The Lesson of Causal Discovery Algorithms for Quantum
  Correlations: Causal Explanations of Bell-inequality Violations Require
  Fine-Tuning}''.
\newblock \href{https://dx.doi.org/10.1088/1367-2630/17/3/033002}{New Journal
  of Physics {\bf 17}, 033002}~(2015).
\newblock  \href{http://arxiv.org/abs/1208.4119}{arXiv:1208.4119}.

\bibitem{causality_pearl}
Judea Pearl.
\newblock ``{Causality: Models, Reasoning, and Inference}''.
\newblock \href{https://dx.doi.org/10.1017/S0266466603004109}{Cambridge
  University Press}. ~(2009).
\newblock 2nd edition.

\bibitem{Bell_1964}
J.~S. Bell.
\newblock ``{On the Einstein Podolsky Rosen paradox}''.
\newblock \href{https://dx.doi.org/10.1103/PhysicsPhysiqueFizika.1.195}{Physics
  Physique Fizika {\bf 1}, 195--200}~(1964).

\bibitem{Bell_1976}
J.~S. Bell.
\newblock ``{The Theory of Local Beables}''.
\newblock Epistemological Letters {\bf 9}, 11--24~(1976).
\newblock
  url:~\href{{https://doi.org/10.1142/9789812795854\newline\_0078}}{{https://doi.org/10.1142/9789812795854\newline\_0078}}.

\bibitem{cavalcantiClassical2018}
Eric~G. Cavalcanti.
\newblock ``{Classical Causal Models for Bell and Kochen-Specker Inequality
  Violations Require Fine-Tuning}''.
\newblock \href{https://dx.doi.org/10.1103/PhysRevX.8.021018}{Physical Review X
  {\bf 8}, 021018}~(2018).
\newblock  \href{http://arxiv.org/abs/1705.0596}{arXiv:1705.05961}.

\bibitem{Pearl_2021}
J.~C. Pearl and E.~G. Cavalcanti.
\newblock ``{Classical causal models cannot faithfully explain Bell nonlocality
  or Kochen-Specker contextuality in arbitrary scenarios}''.
\newblock \href{https://dx.doi.org/10.22331/q-2021-08-05-518}{Quantum {\bf 5},
  518}~(2021).

\bibitem{Fritz_2012}
Tobias Fritz.
\newblock ``{Beyond Bell's theorem: correlation scenarios}''.
\newblock \href{https://dx.doi.org/10.1088/1367-2630/14/10/103001}{New Journal
  of Physics {\bf 14}, 103001}~(2012).
\newblock  \href{http://arxiv.org/abs/1206.5115}{arXiv:1206.5115}.

\bibitem{Triangle_TC}
Thomas~C. Fraser and Elie Wolfe.
\newblock ``{Causal compatibility inequalities admitting quantum violations in
  the triangle structure}''.
\newblock \href{https://dx.doi.org/10.1103/PhysRevA.98.022113}{Physical Review
  A {\bf 98}, 022113}~(2018).
\newblock  \href{http://arxiv.org/abs/1709.0624}{arXiv:1709.06242}.

\bibitem{bipartipe_CoiteuxRoy}
Xavier Coiteux-Roy, Elie Wolfe, and Marc-Olivier Renou.
\newblock ``{No Bipartite-Nonlocal Causal Theory Can Explain Nature's
  Correlations}''.
\newblock \href{https://dx.doi.org/10.1103/PhysRevLett.127.200401}{Physical
  Review Letters {\bf 127}, 200401}~(2021).
\newblock  \href{http://arxiv.org/abs/2105.0938}{arXiv:2105.09381}.

\bibitem{Lauand2023}
Pedro Lauand, Davide Poderini, Ranieri Nery, George Moreno, Lucas Pollyceno,
  Rafael Rabelo, and Rafael Chaves.
\newblock ``{Witnessing Nonclassicality in a Causal Structure with Three
  Observable Variables}''.
\newblock \href{https://dx.doi.org/10.1103/PRXQuantum.4.020311}{PRX Quantum
  {\bf 4}, 020311}~(2023).
\newblock  \href{http://arxiv.org/abs/2211.1334}{arXiv:2211.13349}.

\bibitem{experimental_instrumental}
Iris Agresti, Davide Poderini, Beatrice Polacchi, Nikolai Miklin, Mariami
  Gachechiladze, Alessia Suprano, Emanuele Polino, Giorgio Milani, Gonzalo
  Carvacho, Rafael Chaves, and Fabio Sciarrino.
\newblock ``{Experimental test of quantum causal influences}''.
\newblock
  \href{https://dx.doi.org/https://doi.org/10.1126/sciadv.abm1515}{Science
  Advances {\bf 8}, eabm1515}~(2022).
\newblock  \href{http://arxiv.org/abs/2108.0892}{arXiv:2108.08926}.

\bibitem{Experimental_Triangle}
Emanuele Polino, Davide Poderini, Giovanni Rodari, Iris Agresti, Alessia
  Suprano, Gonzalo Carvacho, Elie Wolfe, Askery Canabarro, George Moreno,
  Giorgio Milani, Robert~W. Spekkens, Rafael Chaves, and Fabio Sciarrino.
\newblock ``{Experimental nonclassicality in a causal network without assuming
  freedom of choice}''.
\newblock \href{https://dx.doi.org/10.1038/s41467-023-36428-w}{Nature
  Communications {\bf 14}, 909}~(2023).
\newblock  \href{http://arxiv.org/abs/2210.0726}{arXiv:2210.07263}.

\bibitem{Wolfe_inflation}
Elie Wolfe, Robert~W. Spekkens, and Tobias Fritz.
\newblock ``{The Inflation Technique for Causal Inference with Latent
  Variables}''.
\newblock \href{https://dx.doi.org/10.1515/jci-2017-0020}{Journal of Causal
  Inference {\bf 7}, 20170020}~(2019).
\newblock  \href{http://arxiv.org/abs/1609.0067}{arXiv:1609.00672}.

\bibitem{Navascues_inflation_complete}
Miguel Navascu\'{e}s and Elie Wolfe.
\newblock ``{The Inflation Technique Completely Solves the Causal Compatibility
  Problem}''.
\newblock \href{https://dx.doi.org/10.1515/jci-2018-0008}{Journal of Causal
  Inference {\bf 8}, 70--91}~(2020).
\newblock  \href{http://arxiv.org/abs/1707.0647}{arXiv:1707.06476}.

\bibitem{Rosset_bound}
Denis Rosset, Nicolas Gisin, and Elie Wolfe.
\newblock ``{Universal bound on the cardinality of local hidden variables in
  networks}''.
\newblock \href{https://dx.doi.org/10.26421/QIC18.11-12-2}{Quantum Information
  and Computation {\bf 18}, 0910--0926}~(2018).
\newblock  \href{http://arxiv.org/abs/1709.0070}{arXiv:1709.00707}.

\bibitem{Fraser_Combinatorial_Solution}
Thomas Fraser.
\newblock ``{A Combinatorial Solution to Causal Compatibility}''.
\newblock \href{https://dx.doi.org/10.1515/jci-2019-0013}{Journal of Causal
  Inference {\bf 8}, 22--53}~(2020).
\newblock  \href{http://arxiv.org/abs/1902.0709}{arXiv:1902.07091}.

\bibitem{Zjawin_restricted}
Beata Zjawin, Elie Wolfe, and Robert~W. Spekkens.
\newblock ``{Restricted Hidden Cardinality Constraints in Causal
  Models}''~(2021).
\newblock  \href{http://arxiv.org/abs/2109.0565}{arXiv:2109.05656}.

\bibitem{Berkeley_Workshop}
Thomas Richardson and Frederick Eberhart, editors.
\newblock ``{Quantum Physics and Statistical Causal Models Workshop}''.
\newblock Simons Institute for the Theory of Computing~(2022).
\newblock
  url:~\href{{https://simons.berkeley.edu/workshops\newline/quantum-physics-statistical-causal-models}}{{https://simons.berkeley.edu/workshops\newline/quantum-physics-statistical-causal-models}}.

\bibitem{PI_Workshop}
``{Causal Inference \& Quantum Foundations Workshop}''.
\newblock {Perimeter Institute for Theoretical Physics}~(2023).
\newblock  url:~\href{https://pirsa.org/C23017}{pirsa.org/C23017}.

\bibitem{hensonTheoryindependent2014}
Joe Henson, Raymond Lal, and Matthew~F. Pusey.
\newblock ``{Theory-Independent Limits on Correlations from Generalized
  Bayesian Networks}''.
\newblock \href{https://dx.doi.org/10.1088/1367-2630/16/11/113043}{New Journal
  of Physics {\bf 16}, 113043}~(2014).
\newblock  \href{http://arxiv.org/abs/1405.2572}{arXiv:1405.2572}.

\bibitem{barrett2020quantum}
Jonathan Barrett, Robin Lorenz, and Ognyan Oreshkov.
\newblock ``{Quantum Causal Models}''~(2020).
\newblock  \href{http://arxiv.org/abs/1906.10726}{arXiv:1906.10726}.

\bibitem{AllenQuantumCommonCauses2017}
John-Mark~A. Allen, Jonathan Barrett, Dominic~C. Horsman, Ciar\'an~M. Lee, and
  Robert~W. Spekkens.
\newblock ``{Quantum Common Causes and Quantum Causal Models}''.
\newblock \href{https://dx.doi.org/10.1103/PhysRevX.7.031021}{Phys. Rev. X {\bf
  7}, 031021}~(2017).
\newblock  \href{http://arxiv.org/abs/1609.0948}{arXiv:1609.09487}.

\bibitem{costaQuantum2016}
Fabio Costa and Sally Shrapnel.
\newblock ``{Quantum Causal Modelling}''.
\newblock \href{https://dx.doi.org/10.1088/1367-2630/18/6/063032}{New Journal
  of Physics {\bf 18}, 063032}~(2016).
\newblock  \href{http://arxiv.org/abs/1512.0710}{arXiv:1512.07106}.

\bibitem{quantum_inflation}
Elie Wolfe, Alejandro Pozas-Kerstjens, Matan Grinberg, Denis Rosset, Antonio
  Ac\'{i}n, and Miguel Navascu\'es.
\newblock ``{Quantum Inflation: A General Approach to Quantum Causal
  Compatibility}''.
\newblock \href{https://dx.doi.org/10.1103/PhysRevX.11.021043}{Physical Review
  X {\bf 11}, 021043}~(2021).
\newblock  \href{http://arxiv.org/abs/1909.1051}{arXiv:1909.10519}.

\bibitem{Weilenmann2020analysingcausal}
Mirjam Weilenmann and Roger Colbeck.
\newblock ``{Analysing causal structures in generalised probabilistic
  theories}''.
\newblock \href{https://dx.doi.org/10.22331/q-2020-02-27-236}{{Quantum} {\bf
  4}, 236}~(2020).

\bibitem{multipartite_nonlocal}
Xavier Coiteux-Roy, Elie Wolfe, and Marc-Olivier Renou.
\newblock ``{Any physical theory of nature must be boundlessly multipartite
  nonlocal}''.
\newblock \href{https://dx.doi.org/10.1103/PhysRevA.104.052207}{Phys. Rev. A
  {\bf 104}, 052207}~(2021).

\bibitem{chaves_informationtheoretic_2015}
Rafael Chaves, Christian Majenz, and David Gross.
\newblock ``{Information-theoretic implications of quantum causal
  structures}''.
\newblock \href{https://dx.doi.org/10.1038/ncomms6766}{Nature Communications
  {\bf 6}, 5766}~(2015).

\bibitem{forre2017markov}
Patrick Forr\'{e} and Joris~M. Mooij.
\newblock ``{Markov Properties for Graphical Models with Cycles and Latent
  Variables}''~(2017).
\newblock  \href{http://arxiv.org/abs/1710.08775}{arXiv:1710.08775}.

\bibitem{Bongers2021}
Stephan Bongers, Patrick Forr{\'e}, Jonas Peters, and Joris~M. Mooij.
\newblock ``{Foundations of structural causal models with cycles and latent
  variables}''.
\newblock \href{https://dx.doi.org/10.1214/21-AOS2064}{The Annals of Statistics
  {\bf 49}, 2885 -- 2915}~(2021).

\bibitem{Araujo2017}
Mateus Ara\'ujo, Philippe~Allard Gu\'erin, and \"Amin Baumeler.
\newblock ``{Quantum computation with indefinite causal structures}''.
\newblock \href{https://dx.doi.org/10.1103/PhysRevA.96.052315}{Phys. Rev. A
  {\bf 96}, 052315}~(2017).

\bibitem{Barrett_2021}
Jonathan Barrett, Robin Lorenz, and Ognyan Oreshkov.
\newblock ``{Cyclic quantum causal models}''.
\newblock \href{https://dx.doi.org/10.1038/s41467-020-20456-x}{Nature
  Communications {\bf 12}, 885}~(2021).
\newblock  \href{http://arxiv.org/abs/2002.1215}{arXiv:2002.12157}.

\bibitem{Oreshkov_2012}
Ognyan Oreshkov, Fabio Costa, and {\v{C}}aslav Brukner.
\newblock ``{Quantum correlations with no causal order}''.
\newblock \href{https://dx.doi.org/10.1038/ncomms2076}{Nature Communications
  {\bf 3}, 1092}~(2012).
\newblock  \href{http://arxiv.org/abs/1105.4464}{arXiv:1105.4464}.

\bibitem{schmid2023review}
David Schmid, Y{\`i}l{\`e} Y{\=\i}ng, and Matthew Leifer.
\newblock ``{A review and analysis of six extended Wigner's friend
  arguments}''~(2023).
\newblock  \href{http://arxiv.org/abs/2308.16220}{arXiv:2308.16220}.

\bibitem{bongStrong2020}
Kok-Wei Bong, An\'{\i}bal Utreras-Alarc\'{o}n, Farzad Ghafari, Yeong-Cherng
  Liang, Nora Tischler, Eric~G. Cavalcanti, Geoff~J. Pryde, and Howard~M.
  Wiseman.
\newblock ``{A Strong No-Go Theorem on the Wigner's Friend Paradox}''.
\newblock \href{https://dx.doi.org/10.1038/s41567-020-0990-x}{Nature Physics
  {\bf 16}, 1199--1205}~(2020).
\newblock  \href{http://arxiv.org/abs/1907.0560}{arXiv:1907.05607}.

\bibitem{wisemanThoughtful2022}
Howard~M. Wiseman, Eric~G. Cavalcanti, and Eleanor~G. Rieffel.
\newblock ``A ``thoughtful'' local friendliness no-go theorem: a prospective
  experiment with new assumptions to suit''.
\newblock \href{https://dx.doi.org/10.22331/q-2023-09-14-1112}{Quantum {\bf 7},
  1112}~(2023).

\bibitem{wisemanCausarum2017}
Howard~M. Wiseman and Eric~G. Cavalcanti.
\newblock ``{Causarum Investigatio and the Two Bell's Theorems of John Bell}''.
\newblock In Reinhold Bertlmann and Anton Zeilinger, editors, Quantum
  [{{Un}}]{{Speakables II}}: {{Half}} a {{Century}} of {{Bell}}'s {{Theorem}}.
\newblock \href{https://dx.doi.org/10.1007/978-3-319-38987-5\_6}{Pages
  119--142}.
\newblock The {{Frontiers Collection}}. {Springer International
  Publishing}~(2017).
\newblock  \href{http://arxiv.org/abs/1503.0641}{arXiv:1503.06413}.

\bibitem{cavalcantiImplications2021}
Eric~G. Cavalcanti and Howard~M. Wiseman.
\newblock ``{Implications of Local Friendliness Violation for Quantum
  Causality}''.
\newblock \href{https://dx.doi.org/10.3390/e23080925}{Entropy {\bf 23},
  925}~(2021).
\newblock  \href{http://arxiv.org/abs/2106.0406}{arXiv:2106.04065}.

\bibitem{proietti2019experimental}
Massimiliano Proietti, Alexander Pickston, Francesco Graffitti, Peter Barrow,
  Dmytro Kundys, Cyril Branciard, Martin Ringbauer, and Alessandro Fedrizzi.
\newblock ``{Experimental Test of Local Observer Independence}''.
\newblock \href{https://dx.doi.org/10.1126/sciadv.aaw9832}{Science Advances
  {\bf 5}, eaaw9832}~(2019).
\newblock  \href{http://arxiv.org/abs/1902.05080}{arXiv:1902.05080}.

\bibitem{greseleCausal2022}
Luigi Gresele, Julius {von K{\"u}gelgen}, Jonas~M. K{\"u}bler, Elke Kirschbaum,
  Bernhard Sch{\"o}lkopf, and Dominik Janzing.
\newblock ``{Causal Inference Through the Structural Causal Marginal
  Problem}''~(2022).
\newblock  \href{http://arxiv.org/abs/2202.0130}{arXiv:2202.01300}.

\bibitem{triantafillouLearning2010}
Sofia Triantafillou, Ioannis Tsamardinos, and Ioannis Tollis.
\newblock ``{Learning Causal Structure from Overlapping Variable Sets}''.
\newblock In Proceedings of the {{Thirteenth International Conference}} on
  {{Artificial Intelligence}} and {{Statistics}}.
\newblock
  \href{https://dx.doi.org/https://doi.org/10.1007/3-540-36182-0_17}{Pages
  860--867}.
\newblock {JMLR Workshop and Conference Proceedings}~(2010).

\bibitem{vorobev_consistent_1962}
N.~N. Vorob'ev.
\newblock ``{Consistent Families of Measures and Their Extensions}''.
\newblock \href{https://dx.doi.org/10.1137/1107014}{Theory of Probability \&
  Its Applications {\bf 7}, 147--163}~(1962).

\bibitem{augusiak2014elemental}
R.~Augusiak, M.~Demianowicz, M.~Paw{\l}owski, J.~Tura, and A.~Ac{\'i}n.
\newblock ``{Elemental and Tight Monogamy Relations in Nonsignalling
  Theories}''.
\newblock \href{https://dx.doi.org/10.1103/PhysRevA.90.052323}{Physical Review
  A {\bf 90}, 052323}~(2014).
\newblock  \href{http://arxiv.org/abs/1307.6390}{arXiv:1307.6390}.

\bibitem{moreno2022events}
George Moreno, Ranieri Nery, Cristhiano Duarte, and Rafael Chaves.
\newblock ``{Events in Quantum Mechanics Are Maximally Non-Absolute}''.
\newblock \href{https://dx.doi.org/10.22331/q-2022-08-24-785}{Quantum {\bf 6},
  785}~(2022).

\bibitem{adlam2023does}
Emily Adlam.
\newblock ``{What Does `(Non)-Absoluteness of Observed Events' Mean?}''~(2023).
\newblock  \href{http://arxiv.org/abs/2309.03171}{arXiv:2309.03171}.

\bibitem{Geiger1988}
D.~Geiger.
\newblock ``{Towards the formalization of informational dependencies}''.
\newblock \href{https://dx.doi.org/}{UCLA Computer Science}. ~(1988).

\bibitem{verma_pearl}
Thomas Verma and Judea Pearl.
\newblock ``{Causal Networks: Semantics and Expressiveness}''.
\newblock In Ross~D. Shachter, Tod~S. Levitt, Laveen~N. Kanal, and John~F.
  Lemmer, editors, Uncertainty in Artificial Intelligence.
\newblock \href{https://dx.doi.org/10.1016/B978-0-444-88650-7.50011-1}{Volume~9
  of Machine Intelligence and Pattern Recognition, pages 69--76}.
\newblock {North-Holland}~(1990).
\newblock  \href{http://arxiv.org/abs/1304.2379}{arXiv:1304.2379}.

\bibitem{privateMarc}
private correspondence with Marc-Olivier Renou and Victor Gitton.

\bibitem{shpitserIntroduction2014}
Ilya Shpitser, Robin~J. Evans, Thomas~S. Richardson, and James~M. Robins.
\newblock ``{Introduction to Nested Markov Models}''.
\newblock \href{https://dx.doi.org/10.2333/bhmk.41.3}{Behaviormetrika {\bf 41},
  3--39}~(2014).

\bibitem{Danks2002}
David Danks.
\newblock ``{Learning the Causal Structure of Overlapping Variable Sets}''.
\newblock In Steffen Lange, Ken Satoh, and Carl~H. Smith, editors, Discovery
  Science.
\newblock
  \href{https://dx.doi.org/https://doi.org/10.1007/3-540-36182-0_17}{Pages
  178--191}.
\newblock Berlin, Heidelberg~(2002). Springer Berlin Heidelberg.

\bibitem{barrett2006information}
Jonathan Barrett.
\newblock ``{Information processing in generalized probabilistic
  theories}''~(2006).
\newblock  \href{http://arxiv.org/abs/quant-ph/}{arXiv:quant-ph/050821}.

\bibitem{Baumeler_2014}
Amin Baumeler and Stefan Wolf.
\newblock ``{Perfect signaling among three parties violating predefined causal
  order}''.
\newblock In 2014 {IEEE} International Symposium on Information Theory.
\newblock {IEEE}~(2014).

\bibitem{Carla2023}
Carla~A. Ferradini, V.~Vilasini, and V.~Gitton.
\newblock ``{A causal modelling framework for classical and quantum cyclic
  causal structures}''.
\newblock upcoming~(2024).

\bibitem{Vilasini_2022}
V.~Vilasini and Roger Colbeck.
\newblock ``{General framework for cyclic and fine-tuned causal models and
  their compatibility with space-time}''.
\newblock \href{https://dx.doi.org/10.1103/physreva.106.032204}{Physical Review
  A {\bf 106}, 032204}~(2022).
\newblock  \href{http://arxiv.org/abs/2109.1212}{arXiv:2109.12128}.

\bibitem{deutschQuantum1985}
David Deutsch.
\newblock ``{Quantum Theory as a Universal Physical Theory}''.
\newblock \href{https://dx.doi.org/10.1007/BF00670071}{International Journal of
  Theoretical Physics {\bf 24}, 1--41}~(1985).

\bibitem{baumann2019wigners}
Veronika Baumann and {\v{C}}aslav Brukner.
\newblock ``{Wigner's Friend as a Rational Agent}''.
\newblock In Meir Hemmo and Orly Shenker, editors, Quantum, Probability, Logic:
  The Work and Influence of Itamar Pitowsky.
\newblock \href{https://dx.doi.org/10.1007/978-3-030-34316-3\_4}{Pages 91--99}.
\newblock Springer International Publishing~(2020).
\newblock  \href{http://arxiv.org/abs/1901.1127}{arXiv:1901.11274}.

\bibitem{baumann2023observers}
Veronika Baumann and Caslav Brukner.
\newblock ``{Observers in superposition and the no-signaling
  principle}''~(2023).
\newblock  \href{http://arxiv.org/abs/2305.15497}{arXiv:2305.15497}.

\bibitem{liIntroduction2019}
Ming Li and Paul Vit{\'a}nyi.
\newblock ``An {{Introduction}} to {{Kolmogorov Complexity}} and {{Its
  Applications}}''.
\newblock \href{https://dx.doi.org/10.1007/978-3-030-11298-1}{Texts in
  {{Computer Science}}}. Springer International Publishing. Cham~(2019).

\bibitem{rovelli2021relational}
Carlo Rovelli.
\newblock ``{The Relational Interpretation}''.
\newblock In Olival Freire, editor, {The Oxford Handbook of the History of
  Quantum Interpretations}.
\newblock
  \href{https://dx.doi.org/10.1093/oxfordhb/9780198844495.013.0044}{Chapter~43}.
\newblock Oxford University Press~(2022).
\newblock  \href{http://arxiv.org/abs/2109.0917}{arXiv:2109.09170}.

\bibitem{dibiagioStable2021}
Andrea Di~Biagio and Carlo Rovelli.
\newblock ``{Stable Facts, Relative Facts}''.
\newblock \href{https://dx.doi.org/10.1007/s10701-021-00429-w}{Foundations of
  Physics {\bf 51}, 1--13}~(2021).
\newblock  \href{http://arxiv.org/abs/2006.1554}{arXiv:2006.15543}.

\bibitem{fuchs2016qbism}
Christopher~A. Fuchs and Blake~C. Stacey.
\newblock ``{QBism: Quantum Theory as a Hero's Handbook}''~(2016).
\newblock  \href{http://arxiv.org/abs/1612.07308}{arXiv:1612.07308}.

\bibitem{wallace2014emergent}
David Wallace.
\newblock ``{The Emergent Multiverse: Quantum Theory According to the Everett
  Interpretation}''.
\newblock
  \href{https://dx.doi.org/10.1093/acprof:oso/9780199546961.001.0001}{{Oxford
  University Press}}. {Oxford}~(2014).

\bibitem{Frauchiger2018}
Daniela Frauchiger and Renato Renner.
\newblock ``Quantum theory cannot consistently describe the use of itself''.
\newblock \href{https://dx.doi.org/10.1038/s41467-018-05739-8}{Nature
  Communications {\bf 9}, 3711}~(2018).

\bibitem{ormrod2024quantuminfluenceseventrelativity}
Nick Ormrod and Jonathan Barrett.
\newblock ``Quantum influences and event relativity''~(2024).
\newblock  \href{http://arxiv.org/abs/2401.18005}{arXiv:2401.18005}.

\bibitem{vilasini2022generalframeworkconsistentlogical}
V.~Vilasini and Mischa~P. Woods.
\newblock ``A general framework for consistent logical reasoning in wigner's
  friend scenarios: subjective perspectives of agents within a single quantum
  circuit''~(2022).
\newblock  \href{http://arxiv.org/abs/2209.09281}{arXiv:2209.09281}.

\bibitem{ChiribellaProbabilistic2010}
Giulio Chiribella, Giacomo~Mauro D'Ariano, and Paolo Perinotti.
\newblock ``Probabilistic theories with purification''.
\newblock \href{https://dx.doi.org/10.1103/PhysRevA.81.062348}{Phys. Rev. A
  {\bf 81}, 062348}~(2010).

\bibitem{Muller_2021}
Markus M{\"{u}}ller.
\newblock ``Probabilistic theories and reconstructions of quantum theory''.
\newblock \href{https://dx.doi.org/10.21468/scipostphyslectnotes.28}{SciPost
  Physics Lecture Notes}~(2021).

\bibitem{Plavala_2023}
Martin Pl{\'{a}}vala.
\newblock ``General probabilistic theories: An introduction''.
\newblock
  \href{https://dx.doi.org/https://doi.org/10.1016/j.physrep.2023.09.001}{Physics
  Reports {\bf 1033}, 1--64}~(2023).

\bibitem{Geiger1990}
Dan Geiger, Thomas Verma, and Judea Pearl.
\newblock ``{Identifying independence in bayesian networks}''.
\newblock
  \href{https://dx.doi.org/https://doi.org/10.1002/net.3230200504}{Networks
  {\bf 20}, 507--534}~(1990).

\bibitem{Koster99}
Jan T.~A. Koster.
\newblock ``{On the Validity of the Markov Interpretation of Path Diagrams of
  Gaussian Structural Equations Systems with Correlated Errors}''.
\newblock
  \href{https://dx.doi.org/https://doi.org/10.1111/1467-9469.00157}{Scandinavian
  Journal of Statistics {\bf 26}, 413--431}~(1999).

\bibitem{sadeghi2017faithfulness}
Kayvan Sadeghi.
\newblock ``{Faithfulness of Probability Distributions and Graphs}''~(2017).

\bibitem{Brunner2014}
Nicolas Brunner, Daniel Cavalcanti, Stefano Pironio, Valerio Scarani, and
  Stephanie Wehner.
\newblock ``Bell nonlocality''.
\newblock \href{https://dx.doi.org/10.1103/RevModPhys.86.419}{Rev. Mod. Phys.
  {\bf 86}, 419--478}~(2014).

\bibitem{Barrett_2002}
Jonathan Barrett, Daniel Collins, Lucien Hardy, Adrian Kent, and Sandu Popescu.
\newblock ``{Quantum nonlocality, Bell inequalities, and the memory
  loophole}''.
\newblock \href{https://dx.doi.org/10.1103/physreva.66.042111}{Physical Review
  A {\bf 66}, 042111}~(2002).
\newblock  \href{http://arxiv.org/abs/quant-ph/}{arXiv:quant-ph/020501}.

\bibitem{jpa512003bib22}
Emanuel Knill, Scott Glancy, Sae~Woo Nam, Kevin Coakley, and Yanbao Zhang.
\newblock ``{Bell inequalities for continuously emitting sources}''.
\newblock \href{https://dx.doi.org/10.1103/physreva.91.032105}{Physical Review
  A {\bf 91}, 032105}~(2015).
\newblock  \href{http://arxiv.org/abs/1409.7732}{arXiv:1409.7732}.

\bibitem{Bierhorst_2015}
Peter Bierhorst.
\newblock ``{A robust mathematical model for a loophole-free Clauser-Horne
  experiment}''.
\newblock \href{https://dx.doi.org/10.1088/1751-8113/48/19/195302}{Journal of
  Physics A: Mathematical and Theoretical {\bf 48}, 195302}~(2015).
\newblock  \href{http://arxiv.org/abs/1312.2999}{arXiv:1312.2999}.

\end{thebibliography}

\end{document}